\documentclass[twocolumn,nofootinbib,preprintnumbers,amsmath,amssymb]{revtex4}

\usepackage{graphicx}% Include figure files
\usepackage{dcolumn}% Align table columns on decimal point
\usepackage{bm}% bold math
\usepackage{epstopdf}

\begin{document}

\preprint{APS/123-QED}

\title{Detection of the elite structure in a virtual multiplex social system by means of a generalized $K$-core}

\author{Bernat Corominas-Murtra$^1$, Benedikt Fuchs$^1$ and Stefan Thurner$^{1,2,3}$}
%\author{Authors?}

%\thanks{Author correspondence: bernat.corominas@upf.edu\\
%stefan.thurner@meduniwien.ac.at}

\affiliation{
$^1$ Section for Science of Complex Systems; Medical University of Vienna, Spitalgasse 23; A-1090, Austria\\
$^2$Santa Fe Institute, 1399 Hyde Park Road, New Mexico 87501, USA\\
$^3$ IIASA, Schlossplatz  1, A-2361 Laxenburg; Austria}
%\thanks{Author correspondence: bernat.corominas@upf.edu\\

\begin{abstract}
Elites are subgroups of individuals within a society that have the ability and means to influence, lead, govern, and shape societies. Members of elites are often well connected individuals, which enables them to impose their influence to many and to quickly gather, process, and spread information. Here we argue that elites are not only composed of highly connected individuals, but also of intermediaries connecting hubs to form a cohesive and structured elite-subgroup at the core of a social network. For this purpose we present a generalization of the $K$-core algorithm
that allows to identify a social core that is composed of well-connected hubs together with their `connectors'. We show the validity of the idea in the framework of a virtual world defined by a massive multiplayer online game, on which we have complete information of various social networks. 
Exploiting this multiplex structure, we find that the hubs of the generalized $K$-core identify those individuals that are high social performers in terms of a series of indicators that are available in the game. 
In addition, using a combined strategy which involves the generalized $K$-core and the recently introduced $M$-core, the elites of the different 'nations' present in the game are perfectly identified as modules of the generalized $K$-core. Interesting sudden shifts in the composition of the elite cores are observed at deep levels.
We show that elite detection with the traditional $K$-core is not possible in a reliable way. The proposed method might be useful in a series of more general applications, such as community detection.
\end{abstract}

\maketitle

\section*{Introduction}
Almost universally, across cultures and times, societies are structured in a way that a small group of individuals are in the possession of the means to influence, shape, structure, lead, and govern large proportions of entire societies. These selected minorities form the {\em elites}. 
The definition and characterization of an elite is a highly multidimensional and debated problem \cite{Mills:1956, Mills:1958, Keller:1963, Domhoff:1967, Bottomore:1993}. It incorporates considerations about wealth, experience, fame, influence over other individuals, role in societies, clubs, parties, etc. In any case elites can not be defined {\em per se}, but only within the context of a social system, which are superpositions of various time-varying social networks, so-called multiplex networks (MPN) \cite{Mucha:2010, Szell:2010b, Bianconi:2013a}. These networks represent interactions between individuals as links of different types such as communication, trading, friendship, aggression, etc., see Fig. 1a. 
It seems natural that elites have to be defined through their location within these MPNs. 
Indeed, one would generally expect that members of elites are characterized by a large {\em connectivity} \cite{Wasserman:1994} in the various networks of the MPN, which enables them to exert their influence on a large number of other individuals. A large connectivity, paired with a strategic position within the MPN, also allows them to collect, process, and  spread information that is of relevance to them \cite{Freeman:1978}. In this view elites are `core-communities' that, to a certain extent, organise the whole topology of  social interactions in a social system \cite{Wasserman:1994}. 
It is further intuitive that elites are not simply a collection of highly connected individuals, but communities of individuals densely connected (a {\em cohesive subgroup}) containing hubs and maybe other individuals playing functional roles within such elite structure. Moreover, relations among elite members are not incidental: they are defined at the same time at multiple levels, spanning from personal and commercial relationships to information exchanges. The cohesiveness of this group can be achieved by means of direct relations among the elite members or by means of intermediaries, individuals who, although not very connected themselves, establish and coordinate the relations between well connected elite members \cite{Friedkin:1984}. We refer to  these intermediaries as {\em connectors}.

%%%%FIGURE 1%%%%%
\begin{figure}[!ht]
\begin{center}
\includegraphics[width=8.0cm]{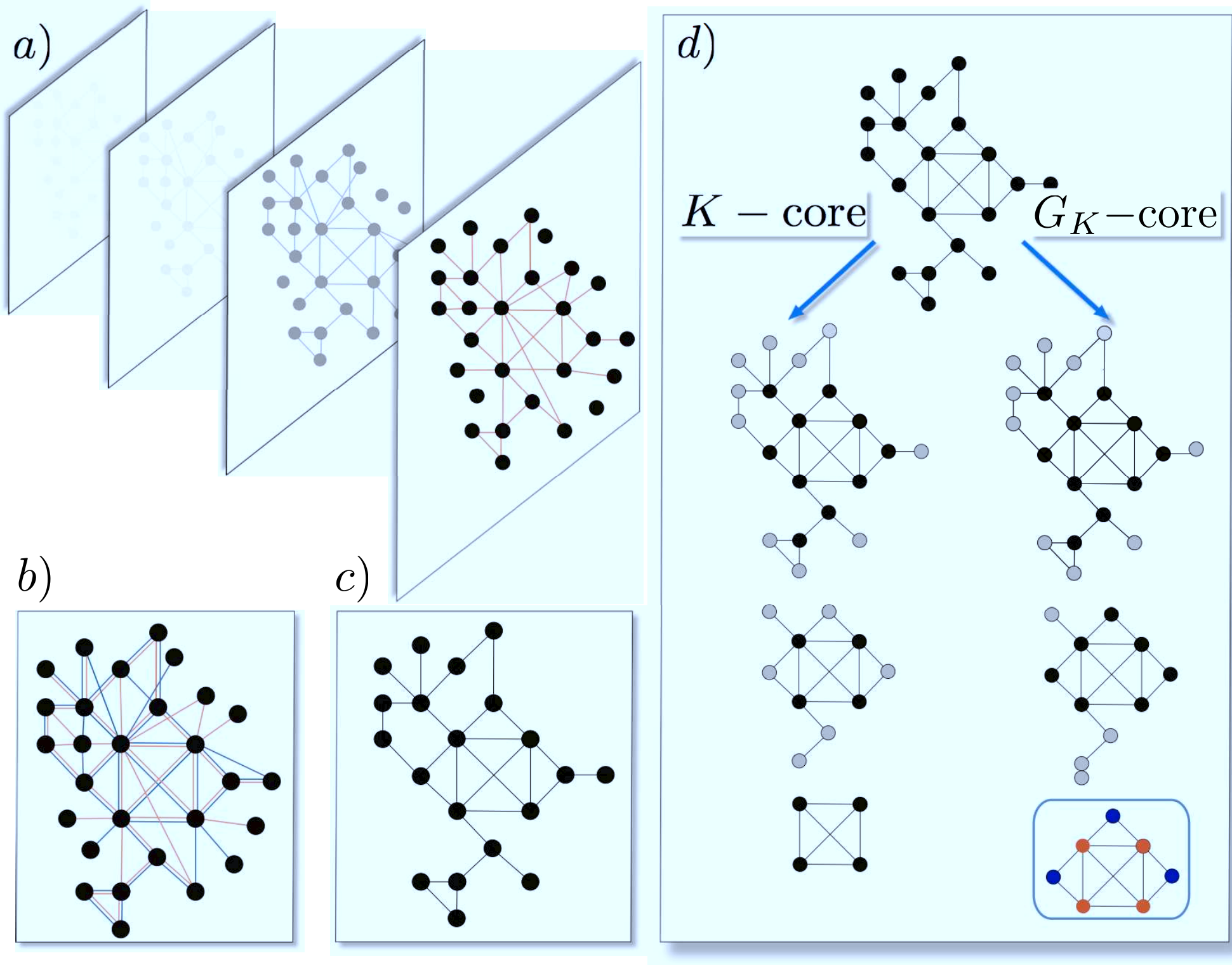}
\caption{{\bf Extracting the core of a Multiplex System}. (a) Representation of multiplex network (MPN) composed of several layers of different relations among nodes. 
(b) A MPN consisting of two link-types {\em orange} and {\em blue}, and (c) its  {\em intersection graph} obtained by keeping those links that are present on both networks. 
(d) Comparison of the $K$-core, left and the {\em generalized} $K$-core, right  algorithms, when applied to the intersection graph: while the $K$-core iteratively removes those nodes whose degree is lower than $K$, (leading to the $K$-core), the $G_K$-core iteratively removes  nodes whose degree is lower than $K$ which are not connected to more than one node whose degree is equal or higher than $K$. We highlight the {\em connectors} (blue) and the hubs (orange). 
Although connectors nodes may have a low degree, they play a role in keeping the overall connectivity at deep levels of network's organization.}
\label{fig:Algorithm}
\end{center}
\end{figure}

%%%%%%%%FIGURE 2%%%%%%
\begin{figure*}%[h]
\begin{center}
\includegraphics[width=17cm]{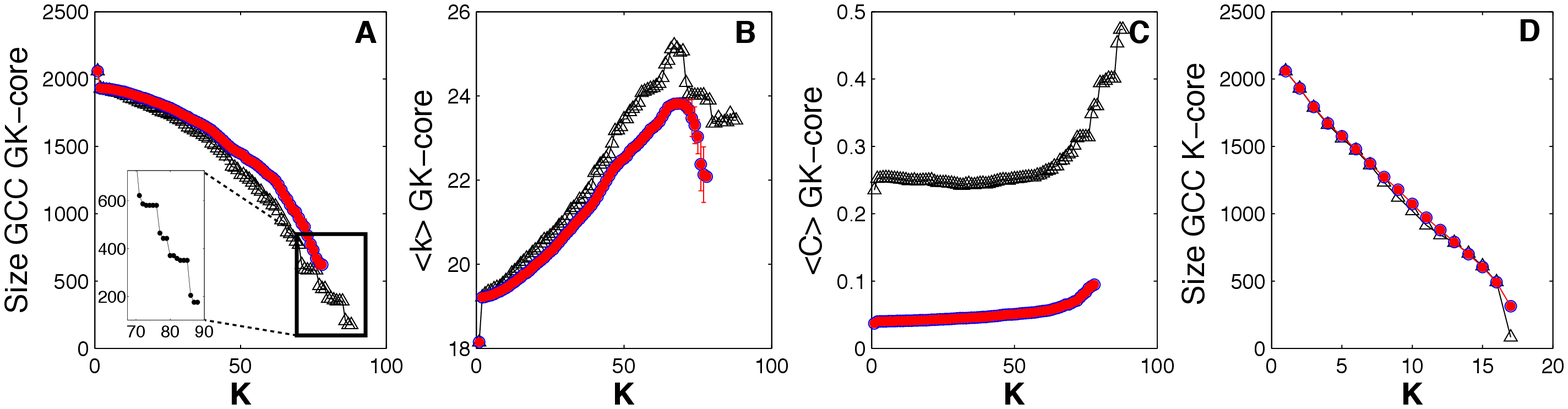}
\caption{{\bf Evolution of the topological indicators along the $G_K$-decomposition sequence for the ${\cal G}_F$ level of the MPN of the period 1140-1200}. In a) we have the evolution of the size of the $GCC$ of the $G_K$-core of the net (black)  and its randomized counterpart (red). In the box inside the figure we highlight the evolution of the size of the $GCC$ of the $G_K$-core at high $K$-levels, where flats regions followed by sudden decreases are observed. b) Evolution of the average degree of the $G_K$-core (black) and its randomized counterpart (red). c) Evolution of the average clustering coefficient of the net (black) against its randomized counterpart (red). Finally, in d) We plot the evolution of the $GCC$ of the $K$-core of the net in terms of $K$ (black) against its randomized counterpart (red). Observe that, for this latter plot, there are no significant statistical differences on the behaviour of the real graph when compared to the randomized one. The results for the random counterpart of the net have been obtained from an ensemble of $25$ randomized versions of ${\cal G}_F$, see text and methods section.}  
\label{fig:TopologicalGF}
\end{center}
\end{figure*}

Given the above considerations, the question arises if one could identify the elite members of a given society from its MPN only by topological means.
The identification of cohesive subgroups at the core of social networks 
has a history of decades and includes the $K$-core decomposition \cite{Seidman:1983, Bollobas:1984, Dorogovtsev:2006}, the clique identification \cite{Harary:1957, Bianconi:2006} or the rich club analysis \cite{Colizza:2006}, among other general methods of cohesive subgroup identification \cite{Girvan:2002, Vicsek:2006}.
In general, these decomposition schemes are focused on the features of the organization of hubs.
However, to adequately describe the organization of a social system, one might think of alternative definitions of `core', taking into account other {\em functional} properties of nodes than just their degree. In the spirit of our definition of elites, connectors should be included in the definition of a core. The heart of this paper is to suggest a generalization of the $K$-core algorithm that naturally takes the `functionality' of connectors into account, and thus allows to detect cores which are composed of hubs together with their connectors. The {\em generalized} $K$-core is obtained by an iterative method inspired both by the so-called $K$-scaffold \cite{Corominas-Murtra:2007, Corominas-Murtra:2008}, and the $K$-core \cite{Seidman:1983,Dorogovtsev:2006}. Specifically, the {\em generalized }$K$-core ($G_K$-core) is the maximal induced subgraph whose nodes either have a degree larger or equal than $K$ or {\em connect} two or more nodes with a degree larger or equal to $K$, see Fig. 1b and methods for details.
We will show that $G_K$-cores isolate the elite communities much more reliably than the traditional $K$-cores. Moreover, as we shall see, $K$-cores and $G_K$-cores show substantial differences in their composition and architecture. 

The quantitative exploration of structural patterns in real social systems is usually hard or even impossible due to poor data availability and due to factors that  escape experimental control. Virtual societies such as those formed in  Massive Multiplayer Online Games (MMOG)  \cite{Castronova:2005} offer an excellent opportunity to avoid these complications and allow for the first time a fully quantitative and empirical understanding of social systems under controlled conditions. Log-files of these games provide complete datasets where practically all actions and interactions of all avatars in the games are recorded.  MMOGs provide a unique framework to test quantitative hypotheses and formulate entirely new questions on social systems. Data then can provide answers at unprecedented levels of precision in the social sciences. 
In this paper we will use data from the MMOG society of the game `Pardus' (http://www.pardus.at) \cite{Szell:2010a}, an open-ended online game with a worldwide player base which currently contains more than 420,000 people. In this game players live in a virtual, futuristic universe where they interact with other players in a multitude of ways to achieve their self-posed goals.  A number of social networks  can be extracted from the Pardus game, leading to the first realization of an entire MPN of a human social system. The MPN consists of the time-varying communication, friendship, trading, enmity, attack, and revenge networks.
These networks are tightly related and mutually influence each other as it has been systematically
explored and quantified in \cite{Szell:2010b, Szell:2010a, Szell:2012a, Szell:2012b, Thurner:2012, Szell:2013, Fuchs:2014}. %For more details on Pardus, see \cite{Pardusgame:2013}.
Here we focus on networks representing {\em cooperative} interactions, namely, {\em friendship} ($F$),  {\em communication} ($C$) and {\em Trade} ($T$). Our social system is therefore given by the MPN  ${\cal M}(t)={\cal M}(V,E_F\times E_C\times E_T,t)$, being $E_F, E_C$ and $E_T$ the sets of links defining a friendship relation, a communicative exchange or a commercial relation, respectively. To ensure the relevance of our results, we will filter the players to rule out the non-active ones. Specifically, we will build the nets over the most active players 'Artemis' universe of the game, which leads us to a set of $~2000-2500$ players.

%\subsection{Core extraction and elite identification}
It is not {\em a priori} clear which link type of the MPN or which combination of links is most relevant for elite detection. A communication link between two individuals might signal an occasional interaction, whereas if a communication link is paired with a trade link, this might be an indication for a much stronger relation between them. For this purpose we derive four more networks, the {\em intersections} among levels of the MPN, see  Fig. 1a,c and methods. 
In these networks a link exists if it is present in two or three of the MPN layers. For these intersection graphs, we formally write ${\cal G}_{FC} = {\cal G}_F\bigcap {\cal G}_C$,  ${\cal G}_{FT} = {\cal G}_F\bigcap {\cal G}_T$,  ${\cal G}_{CT} = {\cal G}_C\bigcap {\cal G}_T$ and ${\cal G}_{FCT} = {\cal G}_F\bigcap{\cal G}_C\bigcap{\cal G}_T$. 
The links of these networks, often called {\em multi-links} \cite{Bianconi:2013b}, encode strong relationships among individuals, for they connect players interacting in more than one type of relation. The strongest links in this sense are those in ${\cal G}_{FCT}$, a graph which we refer to as the structural {\em backbone} of the multiplex system.
The identification of elite structures and core organization is based on the 3 networks of the MPN and their associated four intersection graphs.

The core organization of  ${\cal G}$ will be explored explicitly by computing the sequence of $G_K$-cores, the so-called $G_K$-{decomposition sequence}, which amounts to a `russian doll' decomposition of the networks, 
\[
. . .\subseteq G_K({\cal G})\subseteq G_{K-1}({\cal G})\subseteq . . .\subseteq G_2({\cal G})\subseteq {\cal G}.
\]
The behavior of this sequence of nested levels of networks (either seen in terms of the statistical properties of their graphs, or from their social composition) is  essential to identify the elite organization and the elite structure of our virtual social system. When compared to the traditional $K$-core, we will see that the $G_K$-core provides a much more detailed picture of the nested community structures.
Data from the `Pardus' game enables us to test and compare the quality of the identified core and to see to what extend it relates to properties that are expected for an elite. 
For every player we have a record of  wealth,  leadership role in local organizational structures, and importance in leadership as measured by a `global leadership index'. Local organizational structures are clubs, societies and political parties, in which players organise; we know which player has a leading role in that local organization  which can be president, treasurer  or application master. The global leadership index is a status index that is assigned to each player (visible to all the others) which increases when special tasks (missions) are fulfilled. Such an index is an indicator of the potential influence of the player on decisions affecting the whole `faction' it belongs to. A faction would correspond to a country in the real world. In its current state, the game extends over a universe containing three factions, which are politically independent and lead by their respective elites. 

A final word of caution is needed, in relation to the significance of the data shown here. Since there is no formal/topological definition of elite in a given multiplex society, we adopted the position of showing the averages of the indicators of social relevance of the different core subgraphs we isolate. We checked the position of the topologically isolated sets of nodes within the raw rank of social performance of all players under study. However, an elite is not just a {\em list} of the best performers but a cohesive social structure. therefore, rigorous indicators of statistical relevance would imply the assumption of meaningful null models. This is undoubtedly extremely interesting, but it is an issue going far beyond the scope of this paper. Instead, we adopted the position of giving relevance to our results by confronting them the the ones obtained by means of the K-core, the standard core extraction mechanism, originally designed to extract the network substructure of the most influential individuals in a given society.

% Results and Discussion can be combined.
\section*{Results}
We extract the mentioned seven networks from the Pardus data, in the same way as described in \cite{Szell:2010b, Szell:2010a}. Our analysis is performed over the three networks  ${\cal G}_F, {\cal G}_C$ and ${\cal G}_T$  obtained from the most active players in two time spans of sixty days, $t_1=796-856$ and $t_2=1140-1200$ in units of days since beginning of the game. A link between two players in the layer ${\cal G}_F$ exists if at least one player recognises the other as 'friend' in the whole studied period. Likewise, a link between two players in the layer ${\cal G}_C$ exists if at least one player has sent a message to the other in the studied time span. Finally, a link between two players in ${\cal G}_T$ exists if there has been at least one commercial transaction between these two players within the studied time span. The set of players that will define the set $V$ of the MPN obtained from the period 796-856 contains 2422 players, whereas the set of players defining the MPN of the period 1140-1200 comprises 2059 players. Chosen players are those who are active in at least all three levels of the MPN during all the studied periods. The periods have been chosen using two criteria i) The periods are chosen far away enough from the starting of the game, to ensure that the social structure of the virtual society achieved certain degree of maturity and ii) The comprised time spans do not contain 'war' periods, which may introduce an extra source of noise.

The results of the two time periods under study show a remarkably similar behaviour. Therefore, throughout this section we will mainly show the numerical values of the time period 1140-1200, for the sake of readability. In the supplementary material the reader can find a systematic analysis of the two periods under study.

\subsection*{The backbone exhibits high levels of clustering}
The statistical analysis of networks shows remarkable degree of clustering at all levels of description.
In the period 1140-1200, the average degrees for the various layers of the MPN are $\langle k\rangle_F=18.15$, $\langle k \rangle_C=16.15$,  and $\langle k\rangle_T=33.12$ and the clustering coefficients are remarkably 
high if we take into account these connectivities: $C_F=0.235 (0.037)$, $C_C=0.235 (0.06)$, and  $C_T=0.354(0.04)$. Numbers in brackets correspond to the expected value of the clustering coefficient in an ensemble of random networks  having the same size and degree distribution than the real ones, see methods and   appendix . The intersection networks show a slight decrease on the number of nodes (see Table 1,2 in appendix) and smaller average degrees: $\langle k\rangle_{FC}=6.27$, $\langle k\rangle_{FT}=5.21$, $\langle k\rangle_{TC}=7.05$, and most pronounced, $\langle k \rangle_{FCT}=3.89$, as expected. Although the average degree is lower than in the MPNs, the clustering coefficients still show remarkably high values, especially when compared with the randomized values, $C_{FC}=0.198(0.020)$, $C_{FT}=0.249(0.009)$, $C_{TC}=0.297(0.017)$, and $C_{FCT}=0.197(0.006)$. The  persistence of the clustering coefficient, even for ${\cal G}_{FCT}$, where the expected $C$ for the randomized case almost vanishes, indicates that the mechanism of {\em triadic closure} \cite{Rapoport:1953, Granovetter:1973, Davidsen:2002, Klimek:2013} plays an important role in the dynamical formation of the backbone structure in social systems.

\subsection*{The $G_K$-sequence}
We compute the $G_K$-decomposition sequence (see appendix for details) and observe the following trends. We generally observe long $G_K$- decomposition sequences. The length of the decomposition sequence is the largest value of $K$ for which $G_K$-core is not empty. For the different networks ${\cal G}_{FCT},{\cal G}_{FC},{\cal G}_{FT},{\cal G}_{CT},{\cal G}_F,{\cal G}_C$ and ${\cal G}_T$, these limit values are found at $K=27,38,32,42,88,111$ and, again $111$, respectively.

%%%%%%FIGURE 3%%%%%
\begin{figure*}[!ht]
\begin{center}
\includegraphics[width=16cm]{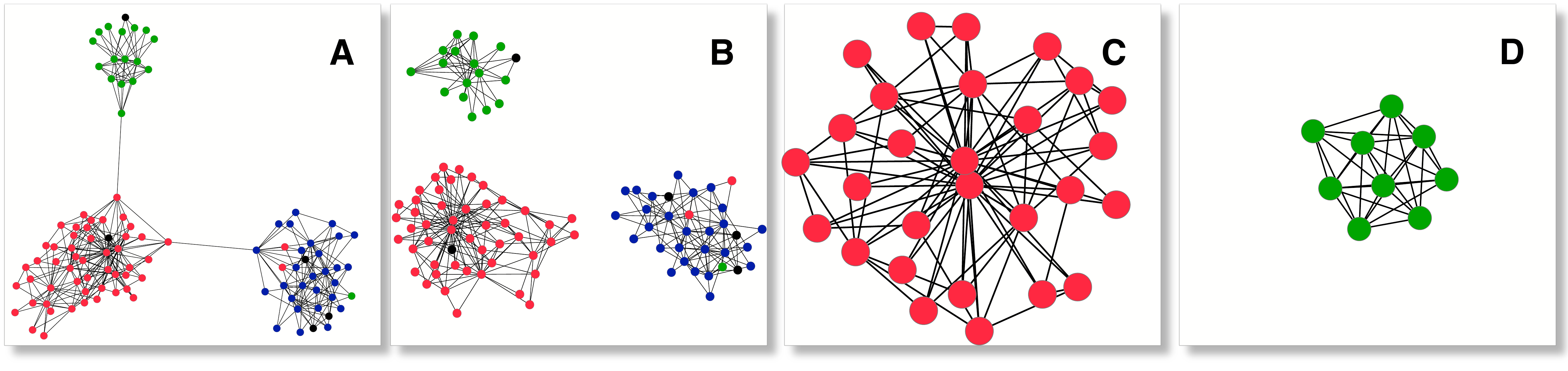}
\caption{{\bf National elites define topological communities at deep levels}. The composition of the $G_K$-core in terms of nations reveals that the multiplex system is organised around the elites of the three existing nations, whose members are depicted with different colours (see text for the use of colours).  We have a) the {\em characteristic }$G_{{K}}$ for ${\cal G}_{FCT}$, b) after the application of the $M$-core ($M=1$), three components appear isolated, to be identified as the three communities composing the $G_K$-core. Such communities are almost uniformly populated by members of the same nation. In c) we have the deepest $G_K$-core, which contains members of only one nation. Interestingly, the composition of the deepest $K$-core, d), is absolutely different from the composition of the deepest $G_K$-core, showing interesting qualitative differences between these two approaches of core extraction. All pictures belong to the period 1140-1200.}
\label{fig:HubsAndLeaders12}
\end{center}
\end{figure*}

In Fig. 2a  the size of the {\em giant connected component}\footnote{In a little abuse of notation, we refer to the $GCC$ as the set of nodes that from a connected component significantly larger than the others, if there exist any.
In our case, the $G_K$-cores generally show a single connected component.} ($GCC$) \cite{Newman:2001} along the $G_K$-decomposition sequence is shown for the ${\cal G}_F$ network (black).
We observe that the $G_K$-decomposition sequence is longer than the one expected by chance, see Fig. 2a, (red).
The situation for the traditional $K$-core is different, with a behaviour similar to the one expected by chance in all studied subgraphs, see Fig. 2d.
Further, the evolution of the size $GCC$ of the $G_K$-cores shows plateaus followed by abrupt changes, which may depict different levels of core organization. 
On closer inspection, we find that often these changes signal the collapse of a cluster, which forms a cohesive community at certain level $K$, and which is completely absent at level $K+1$. The structure of the $G_K$-core just before a  collapse represents one organizational level which is replaced by a deeper one, maybe with different topological and social characteristics. We observe that the length of the decomposition sequence strongly depends on the size of the network, a feature probably due to the power law degree distribution they exhibit. As shown in \cite{Corominas-Murtra:2008} for generic sequences of nested subgraphs, the depth of the decomposition sequence diverges for this kind of networks.

The evolution of the average degree $\langle k\rangle$ along the decomposition sequence for the ${\cal G}_F$ network is seen in Fig. 2b (black).  We find  significant differences between the social networks and their randomized counterparts (red). In most cases one observes that the average degrees along the decomposition sequence first increase with $K$, revealing a phenomenon which resembles the so-called {\em rich club} \cite{Colizza:2006}. Here, elements of the $G_K$-core tend to be more connected among themselves than would be expected by chance. We find an  exception in the ${\cal G}_T$ network where there are no significant differences between the real average degrees and those obtained after randomization. This increasing trend usually peaks and stops at deep levels, followed by a slight decrease at the deepest levels, see Fig. 2b. 
The increase is absent in standard models of random graph like the Erd\"os R\'eny \cite{Newman:2001} and Barab\'asi-Albert \cite{Barabasi:1999} networks, see  Fig. S1 of the appendix . This means that the particular structure of the social network determines the functional form of this curve. Since the randomized ensembles also show an increasing trend of  connectivity through the sequences, see Fig. 2b (red), one might expect that the degree distribution is partially responsible of the observed increase. Furthermore, the presence of high clustering could also be responsible for an additional increase of the connectivity of the cores, thus explaining the deviation from their randomized counterparts. 

Finally, the evolution of the clustering coefficient displays two clearly differentiated regions: At low and medium stages of the decomposition sequence it shows a more or less constant behaviour, followed by an increase at later stages of the sequence. This latter increase may also be the footprint of a rich-club phenomenon in the networks under study. It is worth to observe that along the decomposition sequence, the real values of the clustering coefficient are at least one order of magnitude higher than the expected by chance. In Fig. 2c we display the evolution of the clustering coefficient along the decomposition sequence for the ${\cal G}_F$ network.

\subsection*{Identification of characteristic $K$-levels and core communities through the $M$-core}

In the previous section we pointed out that the evolution of the size of the $G_K$-core throughout the decomposition sequence eventually displays sudden decreases, and that such sharp decays might be related to massive collapses of communities the core. Such change might reveal different levels of core organization. 
How to identify such crucial levels and, therefore, communities inside the $G_K$-core? We assume that the cohesiveness of such communities leads to a high degree of transitivity between them, i.e., that the clustering coefficient inside such communities is exceptionally high. This intuition is supported by the extremely high clustering coefficient values found in the system under study, as we reported above. Moreover, we assume that the degree of transitivity between communities is very low namely, that connections between members of different communities are performed by simple links or by means of connector nodes. Under such defining assumptions of core community, the recently introduced $M$-core \cite{Boguna:2013} plays a crucial role. The $M$-core is the {\em maximally induced subgraph in which each link participates at least in $M$ triangles}. Therefore, the application of the $M$-core with $M=1$, $M=2$ over the $G_K$-cores will remove those links (and maybe some nodes) which do no participate in a highly clustered structure, eventually acting as bridges between communities. The {\em unconnected components that may emerge from the application of the $M$-core  ($M=1,2$) to the $G_K$-core  will be the core communities of our graph at level $K$}, see Fig. 3a,b,  methods section and appendix for a detailed information. For the sake of readability, let us refer to the $M$-core of the $G_K$-core as $M(G_K)$. As long as $K$ increases, the number of components of $M(G_K)$ ($M=1,2$) may fluctuate, thereby identifying different organizational levels within the core of the network. Such fluctuations, if any, will define different levels of core organization. In general, the deepest cores of the networks under study display only a single component, and we will put our focus on the last $K$ by which $M(G_K)$ ($M=1,2$) contains more than a single component. We will refer to this level of organization as the {\em characteristic $K$-level of organization}. It may happen that such a level does not exist, then we will conclude that for this network and under our assumptions, the $G_K$-core does not change dramatically its structure throughout the values of $K$. The rationale behind the definition of this characteristic level is clear: we want to study the structure of the core before the last reorganization, for it may contain many topological and properties absent in the deepest one. As we shall see, this methodology is able to perfectly identify core communities in our system, see Fig. 3a,b. It is worth to emphasise that randomized versions of the nets under study always display a single component and no communities --and, thus, no characteristic $K$-levels-- can be identified.

%In addition, the former application to the $M$-core to the $G_K$ subgraph may be used to identify qualitative changes within the organization of the $G_K$-core as long as $K$ grows. Indeed, if ${\cal G}$ is graph whose $M$-core over $G_K$-core displays more than a single component but whose $M$-core of the $G_K$-core 

%
%We quantify the change of the size of the $G_K$-core along the decomposition sequence by computing the relative decrease of the size of its $GCC$ from level ${K}$ to ${K+1}$, which we call $\Delta f_K$, see methods and   appendix . 
%The size changes are shown in Fig. 2d, where big restructuring events appear as clear peaks. Among them, we study in more detail what happens at the level just before the second last sharp decay. We label this level by ${\tilde K}$. 
%%%%%%FIGURE 4%%%%%%
\begin{figure}%[h]
\begin{center}
\includegraphics[width=8cm]{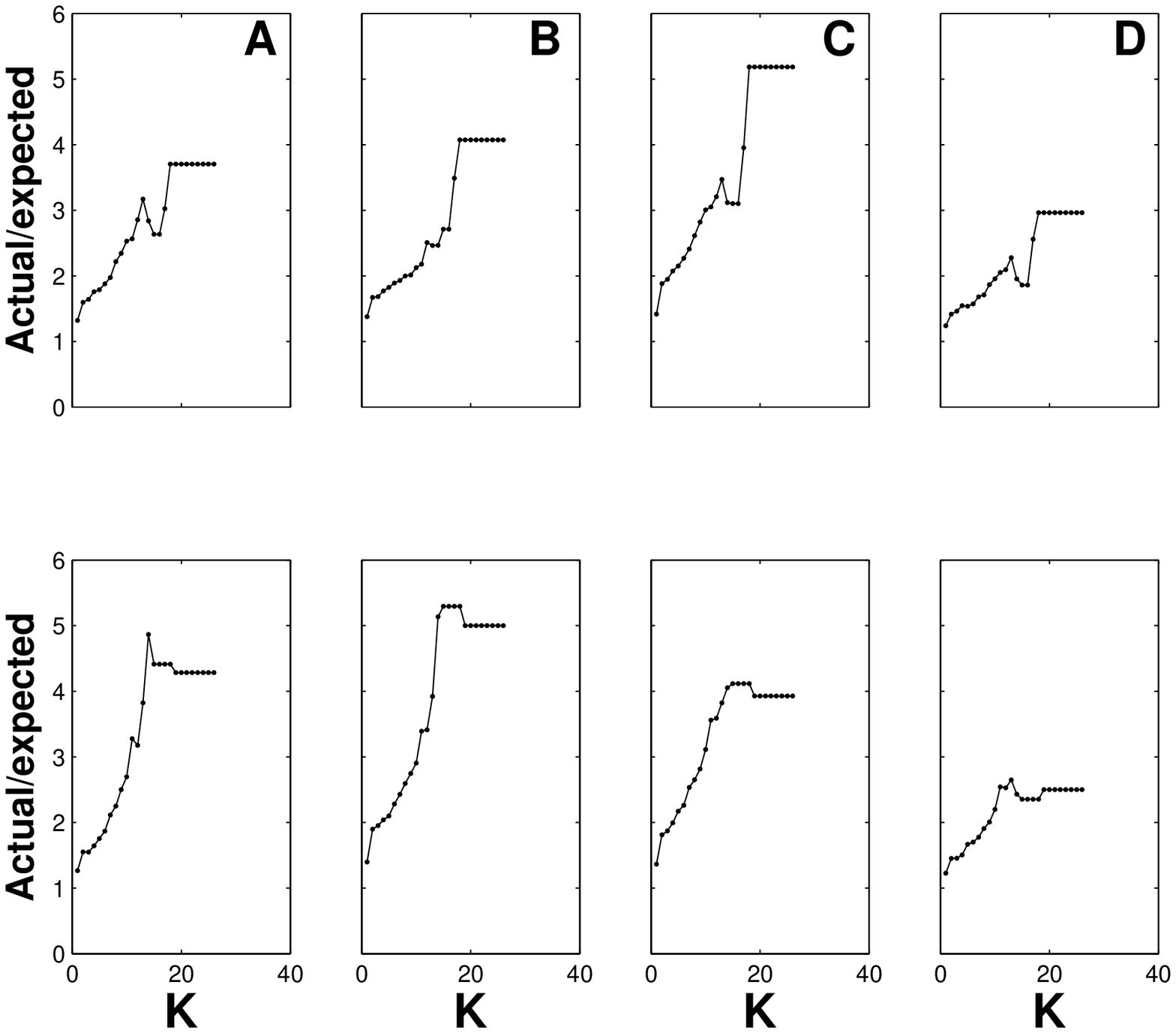}
\caption{{\bf Overabundance of members of the $G_K$-core in the set of the top-$10\%$ best performers of the game}.  In these plots we show the evolution along the $G_K$-decomposition sequence of the quotient between the actual number of members belonging to the $G_K$-core which also belong to the set of the top-$10\%$ best performers of a given indicator against the expected number of them in case they are spread randomly. On top we have the results for the period 756-856 and at the bottom we have the results for the period 1140-1200, both for the ${\cal G}_{FCT}$ networks of their respective periods. We plot this ratio for a) Wealth, b) Global leadership, c) Activity and d) Experience. All of them show an overabundance of members of the $G_K$-core, showing an intrinsic relation between better social performance and deep $G_K$-core membership. It is worth to observe i) the clear overabundance of members of the $G_K$-core within the set of the top $10\%$ in any indicator and ii) the change of the trend after the characteristic $K$-level, which is $K=16$ for the ${\cal G}_{FCT}$ of the period 796-856 and $K=13$ for the period 1140-1200}.
\label{fig:RelTop10}
\end{center}
\end{figure}

%%%%%%%FIGURE 5%%%%%
\begin{figure}%[h]
\begin{center}
\includegraphics[width=8cm]{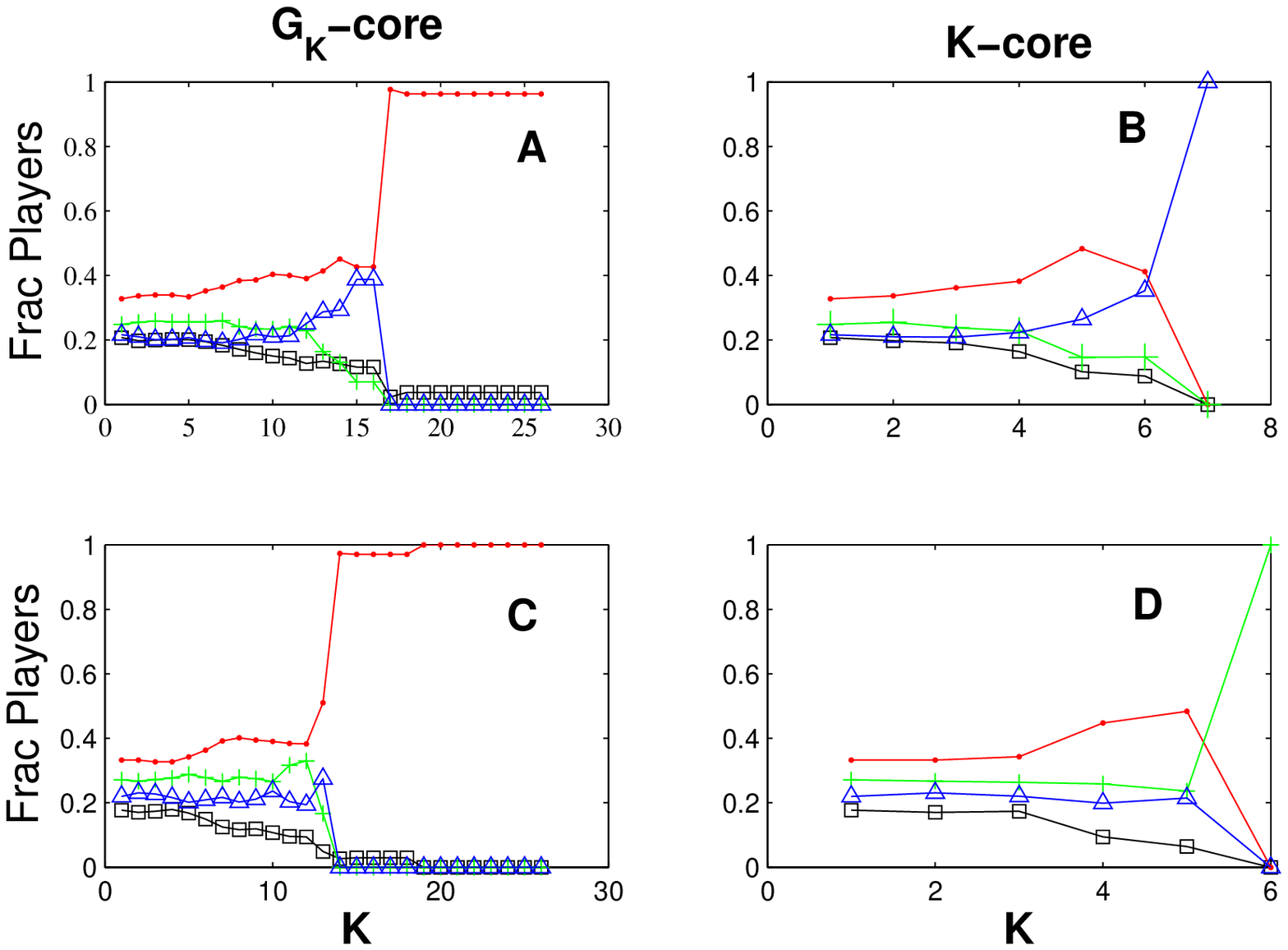}
\caption{{\bf Sharp transitions at the core organization of social networks.} On top a) we have the nation composition of the $G_K$-core and c) the $K$-core  as a function of $K$ for the ${\cal G}_{FCT}$ network corresponding to the period 796-856. At the bottom b) we have the nation composition of the $G_K$-core and (d) the $K$-core in terms of $K$ for the ${\cal G}_{FCT}$ network corresponding to the period 1140-1200. Colours depict the different nations.  As long as $K$ increases, the composition of the cores in terms of nationalities is more or less stationary, with values close to the ones we find in the whole system. At certain $K$ -right after the characteristic $K$- an abrupt change is observed a) for the  and ${\rm b}$), and the composition of the cores becomes uniformly populated by only one nation. The same phenomenon is observed when looking at the $K$-core decomposition sequence, although less pronounced. }
\label{fig:HubsAndLeaders12}
\end{center}
\end{figure}

With the {\em characteristic} $G_{K}$-core and the {\em deepest} $G_{K}$-core, we have two snapshots of the core organization, presumably depicting different structural features. The former represents a core structure which vanishes at deeper levels, the latter shows how the elements at the deepest level of description are organised. 
For the networks corresponding to the period 1140-1200, ${\cal G}_{FCT},{\cal G}_{FC},{\cal G}_{FT},{\cal G}_{CT},{\cal G}_F$, we got the following characteristic $K$-levels: ${K}=13,37,23,38$ and $5$ respectively. ${\cal G}_C$ and ${\cal G}_T$ did not show any characteristic level. The networks obtained out of the intersection of MPN levels display a clearer core community structure and thus relevant characteristic levels can be identified. In the case of ${\cal G}_F$, the characteristic level is found at a very low $K$, so its statistical relevance is lower than the characteristic $K$-levels reported for the intersection nets.

\begin{table*}[!ht]
\caption{{\bf Social indicators of the isolated groups of nodes}. We show the scores for the cores of the ${\cal G}_{FCT}$, ${\cal G}_{FC}$, ${\cal G}_{FT}$and ${\cal G}_{F}$ networks. 'Char. $G_K$' refers to the connectors of the {\em Characteristic} $G_K$, 'Hubs' below it refers to {\em Hubs of the Characteristic $G_K$}.  'Deep. $G_K$' refers to the connectors of the {\em Deepest} $G_K$. 'Hubs' below it refers to {\em Hubs of the Deepest $G_K$}. Deep. $K$-core refers to the nodes of the {\em Deepest $K$-core}. 'All net' refers to all players belonging to the net whose results for the different cores is shown immediately above.We highlighted in boldface the two highest average score for each indicator.}
ÊÊÊÊ\begin{tabular}{c|c|c|c|c|c|c|c|c}
%%%%%%%
ÊÊ& $\langle$Exp$\rangle$ & $\langle$Act$\rangle$ & $\langle$Age$\rangle$ & $\langle$Wealth$\rangle$ & gComp & FracL &$\langle$GlobL$\rangle$ & N  \\ 
\hline
\hline
${\cal G}_{FCT}$ &&&&&&&&\\
\hline 
\hline
Char. $G_K$  & $7.72\times 10^5$ & $5.69\times 10^6$ & $1.02\times 10^3$ & $9.84\times 10^7$ & $0.885$ & $0.195$ & $10.7$ & $87$ \\
Hubs& $1.01\times 10^6$ & $6.86\times 10^6$ & $1.08\times 10^3$ & $1.23\times 10^8$ & $0.933$ & $\mathbf{0.4}$ & $11.4$ & $15 $\\
Deep. $G_K$ & $9.78\times 10^5$ & $5.96\times 10^6$ & $1.09\times 10^3$ & $1.14\times 10^8$ & $0.962$ & $0.154$ & $11.3$ & $26$ \\
Hubs & $5.69\times 10^5$ & $\mathbf{7.39\times 10^6}$ & $\mathbf{1.2\times 10^3}$ & $\mathbf{3.03\times 10^8}$ & $1$ & $\mathbf{1}$ & $12$ & $2$ \\
Deep. $K$-Core  & $7.18\times 10^5$ & $6.23\times 10^6$ & $1.09\times 10^3$ & $\mathbf{1.4\times 10^8}$ & $0.889$ & $0.111$ & $11$ & $9$ \\
All Net & $4.86\times 10^5$ & $3.88\times 10^6$ & $857$ & $4.87\times 10^7$ & $0.875$ & $0.165$ & $7.64$ & $1303$ \\
\hline
\hline
${\cal G}_{FC}$ &&&&&&&&\\
\hline 
\hline
Char. $G_K$ & $8.47\times 10^5$ & $5.72\times 10^6$ & $1.04\times 10^3$ & $7.69\times 10^7$ & $0.884$ & $0.207$ & $9.41$ & $121$ \\
Hubs & $1.32\times 10^6$ & $6.96\times 10^6$ & $\mathbf{1.15\times 10^3}$ & $1.24\times 10^8$ & $0.778$ & $0.333$ & $12.6$ & $9$ \\
Deep.$G_K$& $8.07\times 10^5$ & $5.59\times 10^6$ & $1.01\times 10^3$ & $6.37\times 10^7$ & $0.882$ & $0.235$ & $8.69$ & $85$ \\
Hubs & $\mathbf{1.53\times 10^6}$ & $6.84\times 10^6$ & $1.13\times 10^3$ & $7.26\times 10^7$ & $0.714$ & $0.143$ & $\mathbf{12.7}$ & $7$ \\
Deep. $K$-Core & $9.4\times 10^5$ & $6.03\times 10^6$ & $1.01\times 10^3$ & $6.66\times 10^7$ & $0.882$ & $0.329$ & $9.5$ & $76$ \\
All Net & $4.69\times 10^5$ & $3.72\times 10^6$ & $842$ & $4.35\times 10^7$ & $0.871$ & $0.154$ & $7.4$ & $1600 $\\
\hline
\hline
%%%%%%%%%%%%%%%%%
${\cal G}_{FT}$&&&&&&&&\\
\hline 
\hline
Char. $G_K$ & $8.48\times 10^5$& $5.77\times 10^6$ & $1.05\times 10^3$ & $8.94\times 10^7$ & $0.892$ & $0.169$ & $10.6$ & $83 $\\
Hubs & $\mathbf{1.34\times 10^6}$ & $\mathbf{7.37\times 10^6}$ & $1.13\times 10^3$ & $\mathbf{1.8\times 10^8}$ & $0.889$ & $0.333$ & $12.1$ & $9$ \\
Deep. $G_K$  & $9.2\times 10^5$ & $5.87\times 10^6$ & $1.11\times 10^3$ & $1.1\times 10^8$ & $0.935$ & $0.194$ & $11.3$ & $31$ \\
Hubs & $5.69\times 10^5$ & $\mathbf{7.39\times 10^6}$ & $\mathbf{1.2\times 10^3}$ & $\mathbf{3.03\times 10^8}$ & $1$ & $\mathbf{1}$ & $12$ & $2$ \\
Deep. $K$-Core & $7.18\times 10^5$ & $6.23\times 10^6$ & $1.09\times 10^3$ & $1.4\times 10^8$ & $0.889$ & $0.111$ & $11$ & $9$ \\
All Net & $4.76\times 10^5$ & $3.77\times 10^6$ & $869$ & $4.46\times 10^7$ & $0.872$ & $0.143$ & $7.56$ & $1660$ \\
\hline
\hline
${\cal G}_F$&&&&&&&&\\
\hline 
\hline
Char. $G_K$ & $1.9\times 10^5$ & $1.88\times 10^6$ & $608$ & $1.86\times 10^7$ & $0.86$ & $0.0457$ & $6.08$ & $328$ \\
Hubs & $5.05\times 10^5$ & $4\times 10^6$ & $925$ & $4.65\times 10^7$ & $0.87$ & $0.155$ & $7.61$ & $1585$ \\
Deep. $G_K$ & $7.57\times 10^5$ & $5.34\times 10^6$ & $1.05\times 10^3$ & $5.96\times 10^7$ & $0.877$ & $0.175$ & $7.33$ & $171$ \\
Hubs & $\mathbf{1.39\times 10^6}$ & $6.68\times 10^6$ & $\mathbf{1.15\times 10^3}$ & $7.8\times 10^7$ & $\mathbf{0.6}$ & $0$ & $\mathbf{12.8}$ & $5 $\\
Deep. $K$-Core  & $1.0\times 10^6$ & $6.12\times 10^6$ & $1.08\times 10^3$ & $6.83\times 10^7$ & $0.88$ & $0.253$ & $9.11$ & $83$ \\
%All Net KCore 7 & $4.29\times 10^5$ & $3.47\times 10^6$ & $841$ & $3.95\times 10^7$ & $0.869$ & $0.127$ & $7.23$ & $2058$ \\
%%%%%%%%%%%%%%%%%
\hline
\hline
${\rm All}\;{\rm players}$ 	& $4.3\times 10^5$ 	& $3.5 \times 10^6$ 	& $841$	& $3.96\times 10^7$  &$0.87$    &  $0.12$    & $7.51$   & $2059$
%\hline
ÊÊÊ\end{tabular}
\end{table*}

\subsection*{The $G_K$-core and the elites of the social system}
We can now characterize the individuals populating the cores of the various networks with a series of 
quantitative social indicators in the `Pardus' society. These measure status, competence, social leadership, relevance and success of various kinds. 
In particular we use the following indicators, and we indicate how they appear in Table 1: 
{\em Experience} ($\langle$ Exp$\rangle$, in the table. Numerical indicator accounting for the experience of the player), 
{\em Activity} ($\langle$Act$\rangle$ in the table. Number of actions performed by the player),  
{\em Age} ($\langle$Age$\rangle$ in the table.  Age in units of days after the player joined the game), 
{\em Wealth}, ($\langle$Wealth$\rangle$ numerical indicator accounting for the wealth of the player within the game),
{\em Fraction of leaders} (FracL, in the table. Fraction of players who are leaders in some aspect in a given subgroup of the society at the local level), 
and {\em Global leadership} ($\langle$GlobL$\rangle$ in the table. Numerical indicator evaluating the degree of leadership of the player).
For detailed information about the definition of these indicators, see   appendix. 
We finally checked the {\em gender composition}, the fraction of male/female players in the core. We classify the nodes in the core whether they are a hub or a connector, and present results accordingly. We also computed the scores obtained by the members belonging to the deepest $K$-core, of each studied graph.
In Table 1 we show the scores from four networks ${\cal G}_{FCT}, {\cal G}_{FC}$, ${\cal G}_{FT}$ and ${\cal G}_F$, see SI for  Tables  with all social indicators over  core subgraphs obtained from all networks belonging to the two periods under study.

The combination of the filtering provided by the intersection plus the $G_K$-core extraction clearly identifies the structured groups of players having the highest indicators of social performance and influence. Although, as we pointed out above, there is no null model for an elite detection, one can analyse how relevant are the nodes of the topologically isolated graphs within the collection of raw values of performance indicators belonging to all players of our MPN. Indeed, let us rank all players of the MPN with respect to their performance in a given indicator and then take the 10$\%$ best performers of such indicator. Then, to check if the nodes of our subgraphs are among the best performers we compare the actual number of members which belong both to a given $G_K$-core and to this top-$10\%$ set of players against the expected number of players belonging to the $G_K$-core who also belong to this top-$10\%$ set. What we observe is that, both for wealth and global leadership, the actual number of players of a given $G_K$-core which belong to the set of top-$10\%$ best performers scales up to 5 times the expected one, which shows that there is a strong relation between good performance within the society and being member of the $G_K$-core. In Fig. 4 we show the ratio between the actual number of members of the$G_K$-core belonging to the top-$10\%$ against the expected value. We show the evolution of such ratio for the two periods under study for {\em global leadership}, Fig 4a, {\em Wealth}, Fig 4b, {\em Activity}, Fig 4c, and {\em Experience}, Fig 4d. All plots show an increasing trend which stops around the characteristic $K$-level. Beyond this, the trend flattens and becomes stable, due to the very tiny variations suffered by the $G_K$-core at these levels, until it completely collapses.

In table 1 we highlighted in Boldface the two highest average scores for the following sets of nodes: Connectors of the $G_K$-core at the characteristic $K$-level, Hubs of the $G_K$-core at the characteristic level, Connectors of the deepest $G_K$-core, Hubs of the deepest $G_K$-core and the scores of the players of the whole network. We show the results for ${\cal G}_{FCT}, {\cal G}_{FC}, {\cal G}_{FT}$ and ${\cal G}_F$ for the period 1140-1200. In tables 1,2 of the appendix the reader will find an exhaustive analysis of all the nets belonging to the two periods under study.
Interestingly, the highest scores of a given network are not necessarily found at the deepest level of the decomposition sequence,  but are usually found in the identified characteristic $K$-level, as seen in Table 1 in {\em Experience} in ${\cal G}_{FCT}$ and {\em Wealth} in ${\cal G}_{FC}$.  This happens even though the number of players belonging to the characteristic $K$-level is substantially larger than the number of players populating the deepest $G_K$-core.

We finally check if the membership to the connector set of a $G_K$-core implies a distinction with respect to those players whose connectivity patterns are comparable. Specifically, we refer to individuals having the same degree than a given connector but not being members to the connector set of $G_K$.
Suppose that an individual $v_i$ is a connector in the characteristic $K$-level of ${\cal G}_{FCT}$, ($K=13$, for the period 1140-1200)  with a degree in the ${\cal G}_{FCT}$ network of $k_i$. Now take all individuals in ${\cal G}_{FCT}$ whose degree is equal to $k_i$ but who {\em do not} belong to the characteristic $G_K$ of this net.  
We observe that  the relative performance of connectors with respect to those associated non-connectors of same degree is about $20-40\%$ higher, in particular:  $\langle Exp.\rangle_{G_K}/{\langle Exp.\rangle_{{\rm not-}G_K}} \approx 1.42$,  $\langle Act.\rangle_{G_K}/\langle Act.\rangle_{{\rm not-}G_K}\approx 1.3$, $\langle Age\rangle_{G_K}/\langle Age\rangle_{{\rm not-}G_K} \approx 1.2$ and  
$\langle Wealth\rangle_{G_K}/\langle Wealth\rangle_{{\rm not-}G_K} \approx 1.3$.  These results point to the fact that to belong to the $G_K$-core structure increases the chances of having high scores of social performance. In some cases, we observe that the performance of connectors of the deepest $G_{K}$-core is still higher than the one exhibited by the members of the $K$-core, see, for example, $\langle  Exp.\rangle$ for ${\cal G}_{FCT}$ in Table 1 and  appendix.  
Therefore, connectors, although in general they perform worse than hubs in the $G_K$-cores, could constitute a secondary elite, which presumably takes advantage of the knowledge of the underlying net of relations defining the dynamics of the social system.

\subsection*{$G_K$-core clusters identify national elites / sharp reorganization at deep levels}
We finally look at the national composition of the cores. Players usually belong to one of three `factions' existing in the game, which are the equivalent of countries or nations. These nations are labeled as `nation $1$', `nation $2$' and `nation $3$', associated to colours  red, green and blue, respectively,  in Figs. 3 and 5. Players shown in black are not associated to any nation. Over all the population of the {\em Artemis} universe, the fraction of players in each nation is  $0.34$,   $0.27$ and $0.21$, for nations $1-3$, respectively. Players not associated to any nation represent a fraction of $0.13$ of all players. 

Along the $G_K$-decomposition sequence of all studied networks, the nation composition of the $G_K$-cores displays two well differentiated regions. At lower levels of $K$, the national composition of the $G_K$-core is close to the one corresponding to the whole society. At high $K$-levels, $G_K$-cores are populated only by members of a single nation. The shift between these two qualitatively different core organizations is  abrupt, and occurs right after the characteristic ${K}$-level. This behavior can be clearly seen in Fig. 5a,c, where we plot the evolution of the national
 composition of $G_K$-cores along the $G_K$-decomposition sequence of ${\cal G}_{FCT}$ belonging to the two periods under study.
The evolution of the national composition of the $K$-core also show a similar behaviour, although less abrupt and only at the very late stages of the $K$-core-decomposition sequence, see Fig. 5b,d. 

The application of the $M$-core ($M=1$) over the $G_K$-core shows that the elites of the three nations are clearly identified as clusters at the characteristic ${K}$-level. This can be seen in Fig. 3a,b, where we have the $G_K$-core ${\cal G}_{FCT}$ at the characteristic $K$-level and the $M(G_K)$. As we can see, the proposed method combining the $G_K$-core and the $M$-core perfectly identifies three communities belonging to the three existing nations. Interestingly, the cohesion of the entire core structure across nations is assured only by connectors. At deeper $K$-levels,  only members of one nation populate the $G_K$-core, forming a compact cluster with no community differentiation, see Fig. 3c. The deepest $K$-level of the $K$-core is also populated by individuals belonging all of them to the same nation, see Fig. 3d.
It is worth to remark that, against intuition, the national cluster isolated by the deepest $K$-core differs completely from the one isolated by the deepest $G_K$-core. Finally, it is worth to mention that $10$ of the $13$ identified hubs of the characteristic $G_{{K}}$-core of ${\cal G}_{FCT}$ have a specific leadership role, whereas only $1$ of the $9$ members of the deepest $K$-core does.

\section*{Discussion}
The aim of this study was to propose a topological method to detect the elites in a social system.
We define elites not only as the set of highly connected individuals within a society, but as the set of highly connected ones {\em together} with their connectors in a network  whose links depict multiple relations, like personal, communication or trade ones. Those elites are, presumably, strategically located at the core of the multiplex system defined by the society. To identify the elite cores, we suggest an algorithm that is similar in spirit to the traditional  $K$-core, but that leads to entirely different compositions of the resulting core, which we called the  {\em generalized} $K$-core. As a test system we used the human society of players of the MMOG Pardus, which not only provides the networks of various social interactions \cite{ Szell:2010b, Szell:2010a, Szell:2012a, Szell:2012b, Thurner:2012, Szell:2013}, but also contains quantitative information of how individual players perform socially within the society in terms of leadership, wealth, social status among other skills, in which elite members are expected to score exceptionally high. We find that elite structures are formed by hubs connected either directly or through connectors, generally at deep levels of the core (large $K$). Hubs of these  core subsystems display the highest scores on social relevance, and this is especially true for the backbone network and for the networks obtained out of the intersection of two levels of the MPN, specifically, of friendship and communication levels, and of friendship and trade levels. In addition, we could show that connectors within the $G_K$-core perform consistently worse than hubs, however, we collected evidence pointing to the fact that connectors clearly socially outperform individuals (matched for their degree) that are not part of the $G_K$-core. This indicates that connectors could constitute something like a `secondary' elite within the system, taking advantage of the knowledge they have of the underlying network of social relationships. 
In terms of national composition and core community structure, we have seen that a combined strategy including the use of the recently introduced $M$-core and the $G_K$-core clearly detects the clusters belonging to the elites of the three nations present in the game, thereby providing a new tool for community detection focused on the core properties of the net. Reorganization of the national composition of the cores happens in sharp bursts,  rapid changes which are the footprint of  the collapse of clusters within the core from one level $K$ to another. In all performed analysis, it is worth mentioning the low performance of the $K$-core, when compared to the $G_K$-core to identify those leading subsets of individuals. 
We finally point out that, in spite of their low average degree, in all of the studied networks we found a remarkable level of clustering, which we attribute to the process of triadic-closure that seems to be a major driving force in the dynamics of social network formation \cite{Szell:2010b, Rapoport:1953, Granovetter:1973,  Davidsen:2002, Klimek:2013}.

The presented results suggest that the subgraphs isolated by means of the $G_K$-core actually correspond to the way elites interact and define cohesive subgroups. In more general terms, further works could explore the role of connector nodes in terms of information flow within networks or their presumably relevant role when a dynamical process is defined over the network. It is reasonable to think that the combination of both low connectivity and their role of hinge between clusters may provide them a predominant role in terms of dynamic organization within the network.
The proposed method could lead to a wide range of more general applications, such as network visualization or as a  community detection algorithm.

% You may title this section "Methods" or "Models". 
% "Models" is not a valid title for PLoS ONE authors. However, PLoS ONE
% authors may use "Analysis" 
\section*{Materials and Methods}

%\subsection{Randomisation of networks}
{\em Randomisation of networks.-}
Random ensembles of a given network ${\cal G}$ have been obtained after a rewiring process which keeps the degree of each node invariant. For a real network ${\cal G}$, we created $25$ randomized versions by applying the rewiring operation $100$ times the number of links of ${\cal G}$.

%\subsection{Intersection of different levels of the multiplex system}
{\em Intersection of different levels of the multiplex system.-} We formally refer to multiplex networks (MPNs) as ${\cal M}$, and to single graphs as ${\cal G}$.
In a multiplex graph,  ${\cal M}$, the set of nodes $V=\{v_1, . . .  ,v_n\}$ can be connected by different types of relations or links $\mathbf{E}=\{E_{\alpha_1},. . .,E_{\alpha_M}\}$, $E_{\alpha_k}=\{e_i(\alpha_k),. . .,e_m(\alpha_k)\}$. The whole multiplex is thus described by
\begin{equation}
{\cal M}={\cal M}(V,E_{\alpha_1}\times . . .\times E_{\alpha_M}).\nonumber
\label{eq:Multiplex}
\end{equation}
Let $E'=\{E_{\alpha_i},. . . ,E_{\alpha_k}\}$, $E'\subset \mathbf{E}$, be a subset of the overall type of potential relations that can exist between two nodes, thereby redefining the concept of {\em link} as a collection of relations that relate two given nodes, instead of a single type of relation. We define the $E'$-intersection network, ${\cal G}_{E'}$ as
\begin{equation}
{\cal G}_{E'}={\cal G}\left(V, \bigcap_{E_{\alpha_i}\in E'}E_{\alpha_i}\right).\nonumber
\label{eq:partial_intersection}
\end{equation}
In this network, links connect those pairs of nodes which are connected through, at least, links of type $E_{\alpha_i},. . . ,E_{\alpha_k}$.

%\subsection{The generalized $K$-core}  
{\em The generalized $K$-core.-} The {\em generalized} $K$-core subgraph, $G_K({\cal G})$ of a given graph ${\cal G}$ is  the maximal induced subgraph in which every node is either a hub with  a degree  equal or higher than $K$, or a connector that -- regardless of its degree -- connects at least $2$ hubs with degree equal or higher than $K$. 
It can be obtained through a recursive pruning process. Starting with graph ${\cal G}$ we remove all  nodes $v_i\in {\cal G}$ satisfying that: (1) its degree is lower than $K$ {\em and} (2)  at most one of its nearest neighbors has a degree equal or higher than $K$.  
We iteratively apply this operation over a finite graph ${\cal G}$ until no nodes can be pruned, either because the $G_K$-core is empty or because all nodes which survived the iterative pruning mechanism cannot be removed following the above instructions. The graph obtained after this process is the {\em generalized} $K$-core subgraph.
Note that, for any finite graph, there exists a $K^*$ by which even though $G_{K^*}\neq \varnothing$, $(\forall K>K^*)$ $G_K({\cal G})=\varnothing$. We refer to $G_{K^*}({\cal G}$) as the {\em deepest} $G_K$-core of the network ${\cal G}$, see appendix for the algorithm. 

The standard $K$-core is obtained by means of an iterative algorithm like the one shown above. The step of the algorithm consists in removing nodes whose degree is lower than $K$. This is performed iteratively until there are no more nodes to prune, see  appendix.

Finally, the $M$-core is obtained by means of an iterative algorithm like the ones shown above. The step of the algorithm consists in removing {\em links} participating in less than $M$ triangles. Again, this is performed iteratively until there are no more nodes to prune, see  appendix.
%\subsection{Identifying levels of organization at the core}

{\em Identifying levels of organization at the core.-}
The definition of level of organization is based on the presence of highly clustered communities in the $G_K$-core and its eventual collapse when $K$ increases.
Specifically, given a graph ${\cal G}$:
\begin{itemize}
\item
Compute its $G_K$-core
\item
Compute the $M$-core with $M=1$ over the $G_K$-core and check if the subgraph contains more than a single component. If not, compute the $M$-core ($M=2$) over the $G_K$-core and check if it contains more than a single component.
\item
Components of the $M(G_K)$ are the core communities at level $K$ of the $G_K$-core.
\item
If the $M(G_K)$ with $M=1,2$ contains a different number of components than $M(G_{K+1})$ ($M=1,2$), $K$ is a characteristic level of organization.
\end{itemize}
Throughout the paper we have been focused on the characteristic level of organization defined by the {\em largest $K$ by which $M(G_K)$, ($M=1,2$) contains more than single component.} At deep levels, all the studied $M(G_K)$'s contain only a single component. Furthermore, it may happen that $G_K$ itself contains more than a single component. This does not change the algorithm for characteristic $K$-level identification.

\section*{Acknowledgments}
The authors acknowledge two anonymous reviewers for their comments.
This work was supported by the Austrian Science Fund FWF under KPP23378FW, the EU   LASAGNE project, no. 318132 and the EU MULTIPLEX project, no. 318132. BC-M thanks Andreu Corominas-Murtra for suggesting discussions.

%\newpage
\appendix

\section{Extracting the core}
\subsection{Intersection of different levels of the multiplex system}
Let us have a graph ${\cal G}(V, E)$  where $V=\{v_1, . . .  ,v_n\}$ is the set of {\em nodes} and $E=\{e_i,. . .,e_m\}$ the set of {\em links} connecting this nodes. Given a node $v_i$ its {\em degree} is the number of first neighbors, or nodes a given node is linked to, to be written as $k(v_i)$. The probability that a randomly choosen has degree $k$ is $p(k)$. The first moment of the degree distribution gives us the {\em average degree} $\langle k\rangle=\sum_kkp(k)$ \cite{Newman:2001}. 
Social systems are better described by means of {\em multiplex graphs}, \cite{Szell:2010b} which can be thought of as different graphs sharing the same set of nodes. In a multiplex graph,  ${\cal M}$, the set of nodes $V=\{v_1, . . .  ,v_n\}$ can be connected by different types of relations or links $\mathbf{E}=\{E_{\alpha_1},. . .,E_{\alpha_M}\}$, $E_{\alpha_k}=\{e_i(\alpha_k),. . .,e_m(\alpha_k)\}$. The whole multiplex system is thus described by:
\begin{equation}
{\cal M}={\cal M}(V,E_{\alpha_1}\times . . .\times E_{\alpha_M}).
\label{eq:Multiplex}
\end{equation}
In these networks, concepts such as degree distribution or average degree are relative to the type of relations (links) we are interested in. Now let $E'=\{E_{\alpha_i},. . . ,E_{\alpha_k}\}$, $E'\subset \mathbf{E}$, be a subset of the overall type of potential relations that can exist between two nodes. We define the $E'$-intersection network, ${\cal G}_{E'}$ as follows:
\begin{equation}
{\cal G}_{E'}={\cal G}\left(V, \bigcap_{E_{\alpha_i}\in E'}E_{\alpha_i}\right)
\label{eq:partial_intersection}
\end{equation}
In this network, links connect those pairs of nodes which are connected, at least, by links of type $E_{\alpha_i},. . . ,E_{\alpha_k}$. Links in ${\cal G}_{E'}$ are called {\em multilinks}.

\subsection{The backbone of the multiplex system} 
A special and particularly interesting case of equation (\ref{eq:partial_intersection}) is the graph of the intersection of all types of relations, ${\cal G}_I$ (to be named ${\cal G}_{FCT}$, in the main text), which depicts the {\em backbone} of the multiplex system depicted by ${\cal M}$, namely:
\begin{equation}
{\cal G}_I={\cal G}\left(V, \bigcap_{1\leq i\leq M}E_{\alpha_i}\right).
\label{eq:Intersec}
\end{equation}
We point out that we have to be careful when choosing the different sets of links $E_{\alpha_1}, . . .,E_{\alpha_M}$, since antagonistic relationships (such as enmity and friendship) can lead to empty intersections. We thereby restrict the definition of the intersection graph when this is performed over {\em compatible} sets of links. 

\subsection{The Generalized $K$-core subgraph} 
\begin{figure*}%[h]
\begin{center}
\includegraphics[width=15cm]{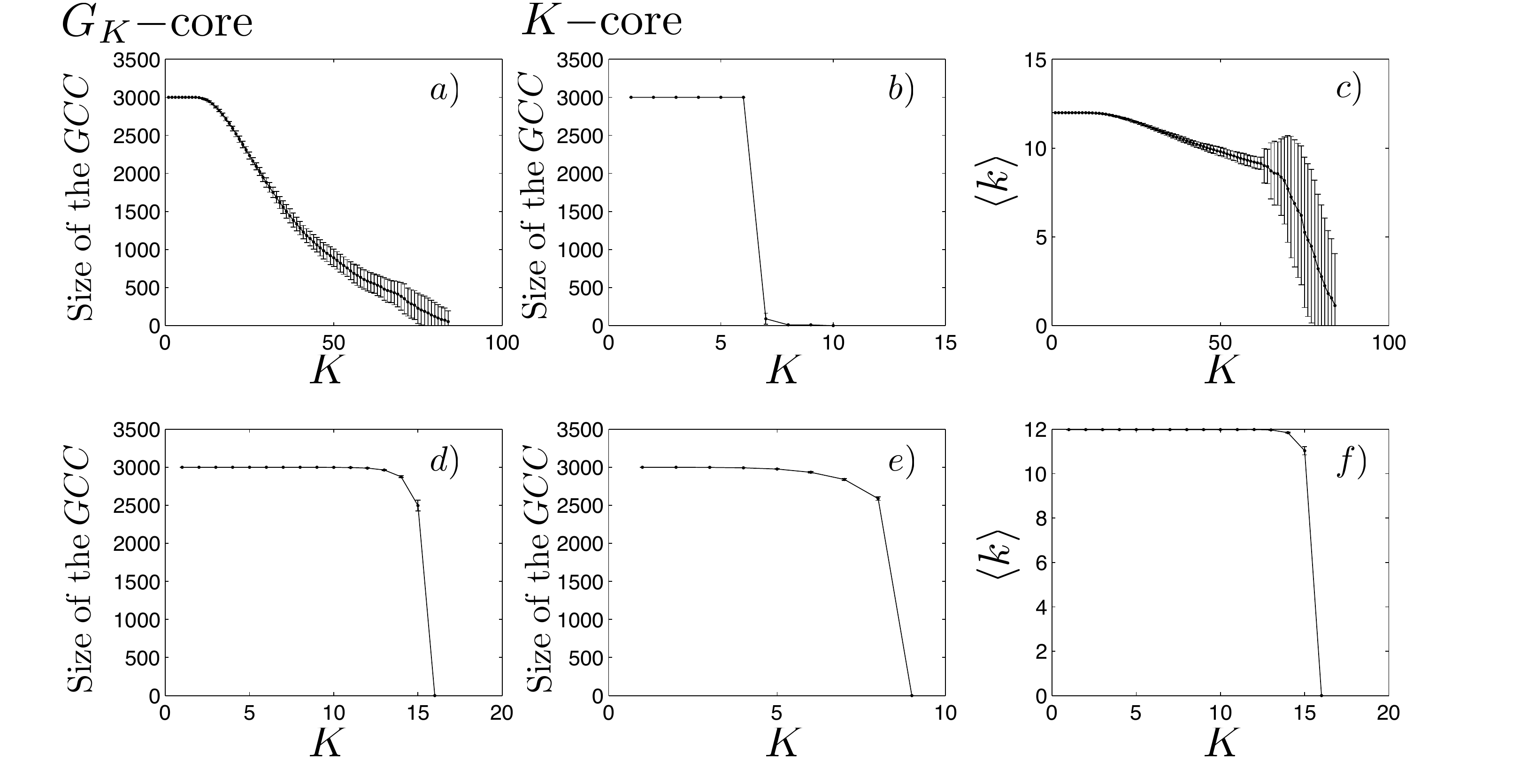}
\caption{Evolution of the size of the $G_K$-cores (a), $K$-cores (b)  and the average connectivities of the $G_K$-cores as a function of the threshold $K$ in a  B-A ensemble containing 3000 nodes and with $\langle k \rangle=12$. Evolution of the size of the $G_K$-cores (d), $K$-cores (e)  and the average connectivities of the $G_K$-cores (f) as a function of the threshold $K$ in a  ER ensemble containing 3000 nodes and with $\langle k \rangle=12$. See text for details.}
\label{fig:NullMods}
\end{center}
\end{figure*}

The {\em Generalized $K$-core}  subgraph of a given graph ${\cal G}$, $G_K({\cal G})$ or $G_K$-core of ${\cal G}$, is  the maximal induced subgraph within which every node is either a hub (its degree is equal or higher than $K$) or a connector (its degree is lower than $K$ but it connects at least $2$ hubs). 
Increasing the threshold $K$ we obtain the {\em decomposition sequence} of ${\cal G}$ in terms of $G_K$, namely:
\[
. . .\subseteq G_K({\cal G})\subseteq G_{K-1}({\cal G})\subseteq . . .\subseteq G_2({\cal G})\subseteq {\cal G}.
\]
We will refer to the above sequence as the $G_K$-{\em decomposition sequence} of ${\cal G}$.
The $G_K$-core of a given graph ${\cal G}$ can be obtained through an iterative pruning process: Suppose an operation $H_K({\cal G})$ by which we prune all the nodes $v_i\in {\cal G}$ satisfying both that  
\begin{itemize}
\item
its degree is lower than $K$ and 
\item
at most $1$ of its nearest neighbors has degree equal or higher than $K$. 
\end{itemize}
If we iteratively apply this operation over a finite graph ${\cal G}$,
\[
H_K^n({\cal G})=\overbrace{H_K\circ . . .\circ H_K}^n(G),
\]
we will reach a value, $n=N$, by which $(\forall M>N)$ $H_K^N({\cal  G})=H_K^M({\cal G})$. We can take it as a definition of the {\em generalized $K$-core}, by saying that:
\begin{equation}
G_K({\cal G})=H_K^N({\cal  G}).
\label{eq:GK}
\end{equation}
The equivalence between this definition and the one provided above can be easily checked: Indeed, on one hand, the algorithm itself forbids the presence of a node which is neither a hub or a connector, because, thank to its iterative nature, it only stops when all {\em surviving} nodes satisfy the conditions to belong to the $G_K$. On the other hand, we observe that the set isolated by the iterative algorithm is maximal: If a node (or a set of nodes) satisfies the conditions imposed by the algorithm, it is not pruned. 
We observe that, in any finite graph, $\exists K^*$ by which although $G_{K^*}\neq \varnothing$, $(\forall K>K^*)$ $G_K({\cal G})=\varnothing$. We will refer to $G_{K^*}({\cal G}$) as the {\em deepest} $G_K$-core of ${\cal G}$. 

Due to the potential richness of connectivity patterns that are allowed inside  the $G_K$-core, we can categorize its nodes according to their topological roles: 
\begin{itemize}
\item
$G^{\rm Con}_{K}({\cal G})$ is the set of nodes of 
$G_K({\cal G})$ whose degree is lower than $K$, the $K$-{\em connectors}, 
\item
$G^{\rm Hub}_{K}({\cal G})$ is the set of nodes of $G_K({\cal G})$ whose degree is equal or higher than $K$, the $K$-{\em hubs}. 
\item
The set of $K$-{\em critical connectors}. A $K$-critical connector is a $K$-connector whose removal implies the breaking of the $G_K({\cal G})$ in two or more parts. We can analogously define the set of $K$-critical hubs. 
\end{itemize}

\subsection{The $K$-core subgraph}
For the $K$-core definition and the exploration of its interesting properties, we refer the interested reader to \cite{Seidman:1983, Bollobas:1984, Dorogovtsev:2006}. The $K$-core of a given graph ${\cal G}$, $KC({\cal G})$, is the maximal induced subgraph whose nodes have degree {\em at least} $K$. It can be obtained through the application of an algorithm qualitatively close to the one described above by iteratively removing nodes whose degree is lower than $K$. The sequence 
\[
. . .\subseteq KC({\cal G})\subseteq {(K-1)}K({\cal G})\subseteq . . .\subseteq 2C({\cal G})\subseteq {\cal G}.
\]
is the $KC$-decomposition sequence, and the largest $K$ by which $KC({\cal G})\neq \varnothing$, the {\em deepest} $K$-core, will be referred to as $K^*C({\cal G})$.

\subsection{The $M$-core subgraph} We end this section by describing the $M$-core subgraph. We refer the interested reader to \cite{Boguna:2013}. The $M$-core of a given graph ${\cal G}$, $M({\cal G})$ is the maximal induced subgraph whose {\em links} participate, {\em at least} in $M$ triangles. The $M$-core can be obtained by the application of an iterative algorithm as the one presented above which iteratively removes links participating in less than $M$ triangles. Although it has not been used for the particular purposes of the current study, we can also define a $M$-core decomposition sequence in the same way we did with the $G_K$-core and the $K$-core.

\section{Model Networks}
We now explore the behavior of the $G_K$ decomposition of two standard models of random graphs, namely, the {\em Erd\"os-R\'enyi} (ER) graph \cite{Newman:2001} and the {\em Barab\'asi-Albert} (BA) graph \cite{Barabasi:1999}. For every type of graph we create an ensemble of $100$ networks each, with $\langle k\rangle=12$ in both the BA and the ER ensemble. We compute the evolution of the {\em Giant Connected Component} of all the non-empty $G_K$-cores and $K$-cores of the corresponding decomposition sequences and we plot them as a function of the threshold defined by $K$, see Fig. \ref{fig:NullMods}.  
For the BA scale-free networks, we observe  a long decomposition sequence, thereby obtaining a picture of the core topology of the net at many different levels, see Fig. \ref{fig:NullMods}a.The behavior of the $G_K$-decomposition sequence for the  ER ensemble shows that the $G_K$-core is either the whole graph or empty which can be due to the almost uniform degree distribution of this kind of graph see Fig. 1d. The behavior of the two ensembles is qualitatively similar under the $KC$-decomposition, showing an all-to-nothing transition at values 
close to $\langle k\rangle/2$, see Fig. 1b,e. The average degree of the successive $G_K$-core subgraphs
shows a slightly descending trend, whereas it remains constant in the case of the ER ensemble, mainly because, if the $G_K$-core is non-empty, it contains almost the whole graph. The counterintuitive decay in $\langle k\rangle$ for the BA ensemble can be explained by the increasing relative abundance of connectors against hubs within $G_K$ as long as $K$ increases.

%\newpage
\section{Indicators of performance}
We explored the behavior of $7$ quantitative indicators of social performance within the `Pardus' game (www.pardus.at):
\begin{itemize}
\item
{\em Experience} is a numerical indicator accounting for the experience of the player, related to battles in which the player has participated, or the number monsters he/she {\em killed}.
\item
{\em Activity} is a numerical indicator related to the number of actions performed by the player.
\item
{\em Age} is the number of days after the player joined the game, 
\item
{\em Wealth}, numerical indicator accounting for the wealth of the player within the game. Wealth accounts for cash money, value of the equipment the player owns within the game.
\item
{\em Fraction of leaders} fraction of players who are leaders in some aspect in a given alliance. Alliance should not be confused with nations. Alliances are small, organized groups of players. In the studied universe, we identify around $\sim 140$ different alliances. Every alliance has its own {\em local} leaders.
\item
{\em Global leadership} is numerical indicator evaluating the degree of leadership of the player. It is increased by doing missions, which are mainly transporting goods or killing monsters.
The higher the {\em Global leadership}, the more powerful items may be bought -- and the more missions are required to reach the next level. In general we can say that the higher this indicator, the more powerful and influential is the player within the whole society defined by the game.
\item
{\em Gender composition} evaluates the fraction of males within a given group of players.
\end{itemize}
In the table we show the scores of all these indicators for all $7$ studied graphs at their respective $\tilde{K}$ and $K^*$-levels. We distinguish between connectors and hubs. We compute theses social indicators for the $K^*$-core as well.c The last column {\em NumberInd} is the number of individuals of the observed subset of nodes.

In tables I and II we provide the above mentioned social indicators of the nodes belonging to i) the critical $G_K$-cores and their hubs, ii) the deepest $G_K$-cores and their hubs, iii) the deepest $K$-core subgraphs and the whole networks. 

\section{Social Networks of the 'Pardus' virtual society}

Throughout the paper we based our analysis in three social networks, namely:
\begin{itemize}
\item Communication network: A link between two players is established if they had a communicative interaction (a player sent a message to the other player, regardless the direction of the informative exchange) within the period under study.
\item Trade network: A link is established if two players had a commercial relation within the period under study.
\item Friendship network: A link is established if a given player identifies the other as 'friend'. This identification an be previous to the period of study; each player has a list of those players who are tagged as friends within the virtual society.
\end{itemize}

We studied two periods of time: i) from day 796 to day 856 and ii) from day 1140 to day 1200; being the day '0' the day in which the game was launched.
In figures 2-8 we study the evolution of basic statistical indicators of the studied networks and the intersections we can extract from them. In black we have the behaviour of the real networks, in read, the average behaviour of an ensemble of 25 randomised versions of the original one. Networks obtained through intersection are randomised {\em after} performing the intersection. Otherwise, it is likely that we would end up quickly to empty networks.
\begin{figure*}%[!ht]
\begin{center}
\includegraphics[width=15.0cm]{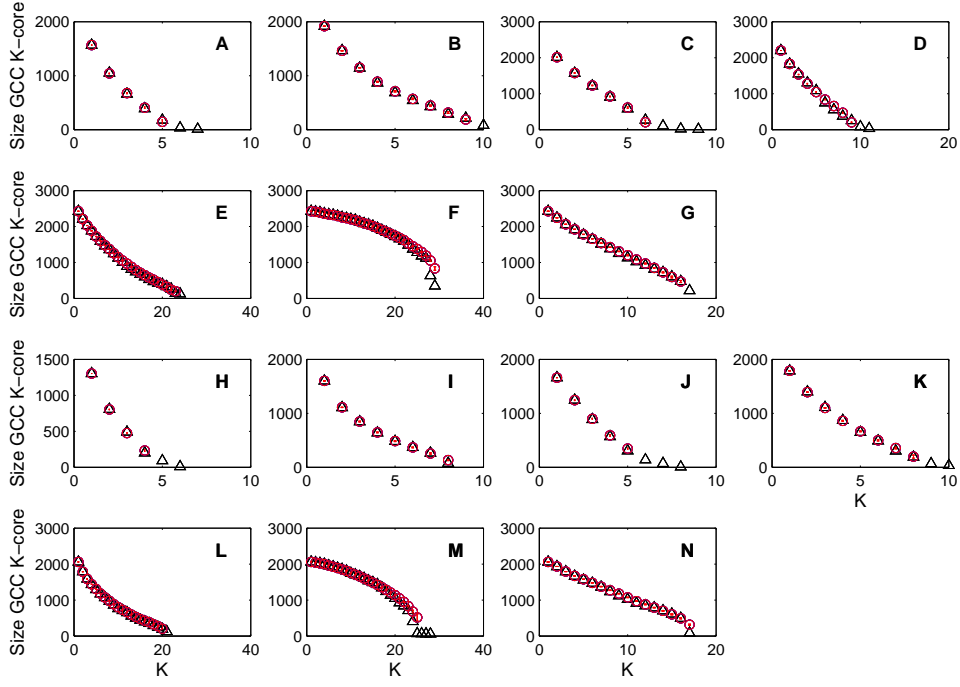}
\caption{Evolution of the size of the  Giant Connected Component of the $K$-core for the networks corresponding to the period 796-856 as a function of $K$. a) ${\cal G}_{FTC}$, b) ${\cal G}_{FC}$, c) ${\cal G}_{FT}$, d) ${\cal G}_{TC}$, e) ${\cal G}_C$, f) ${\cal G}_T$, g) ${\cal G}_{F}$ and the networks corresponding to the period 1140--1200, h) ${\cal G}_{FTC}$, i) ${\cal G}_{FC}$, j) ${\cal G}_{FT}$, k) ${\cal G}_{TC}$, l) ${\cal G}_C$, m) ${\cal G}_T$, n) ${\cal G}_{F}$. Black triangles depict the behaviour of real networks, red circles and their associated error bars depict the average behaviour of an ensemble of 25 randomised versions of the original networks. }
\end{center}
\end{figure*}

\begin{figure*}
\begin{center}
\includegraphics[width=15.0cm]{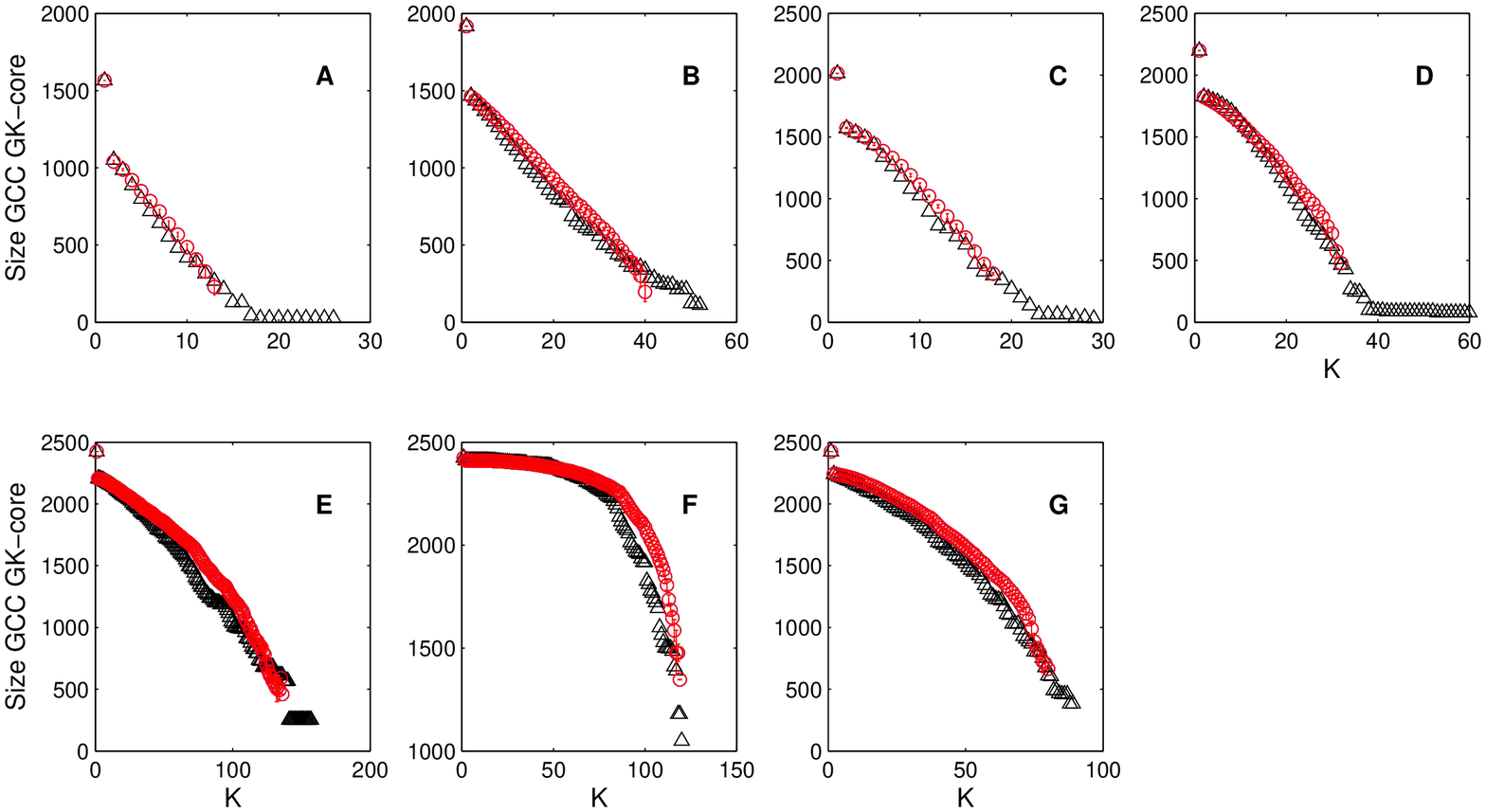}
\caption{Evolution of the size of the Giant Connected Component of the $GK$-core for the networks corresponding to the period 796--856 as a function of $K$. a) ${\cal G}_{FTC}$, b) ${\cal G}_{FC}$, c) ${\cal G}_{FT}$, d) ${\cal G}_{TC}$, e) ${\cal G}_C$, f) ${\cal G}_T$, g) ${\cal G}_{F}$. Black triangles depict the behaviour of real networks, red circles and their associated error bars depict the average behaviour of an ensemble of 25 randomised versions of the original networks.}
\end{center}
\end{figure*}

\begin{figure*}
\begin{center}
\includegraphics[width=15.0cm]{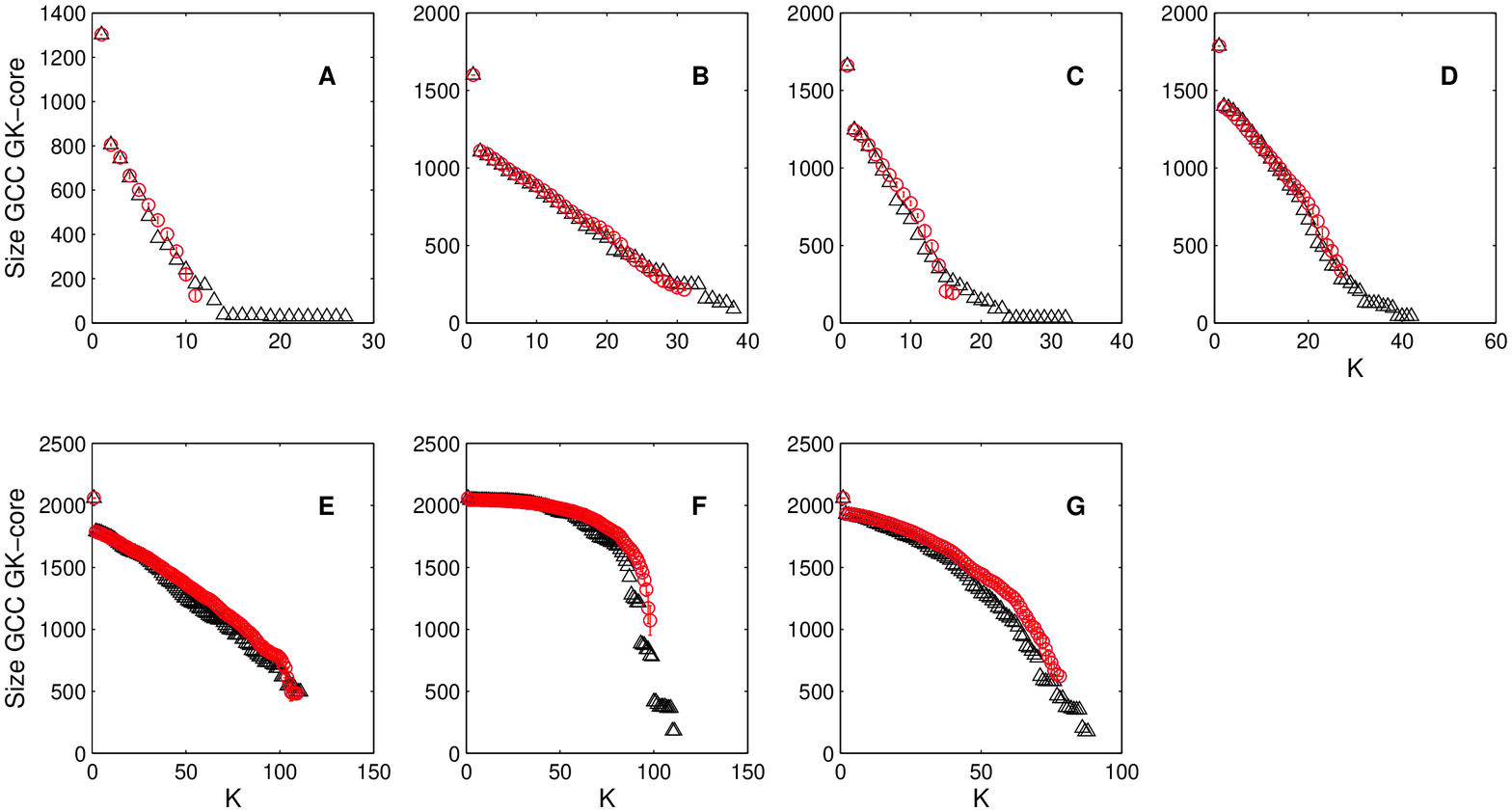}
\caption{Evolution of the size of the Giant Connected Component of the $GK$-core for the networks corresponding to the period 1140--1200 as a function of $K$. a) ${\cal G}_{FTC}$, b) ${\cal G}_{FC}$, c) ${\cal G}_{FT}$, d) ${\cal G}_{TC}$, e) ${\cal G}_C$, f) ${\cal G}_T$, g) ${\cal G}_{F}$. Black triangles depict the behaviour of real networks, red circles and their associated error bars depict the average behaviour of an ensemble of 25 randomised versions of the original networks.}
\end{center}
\end{figure*}

\begin{figure*}
\begin{center}
\includegraphics[width=15.0cm]{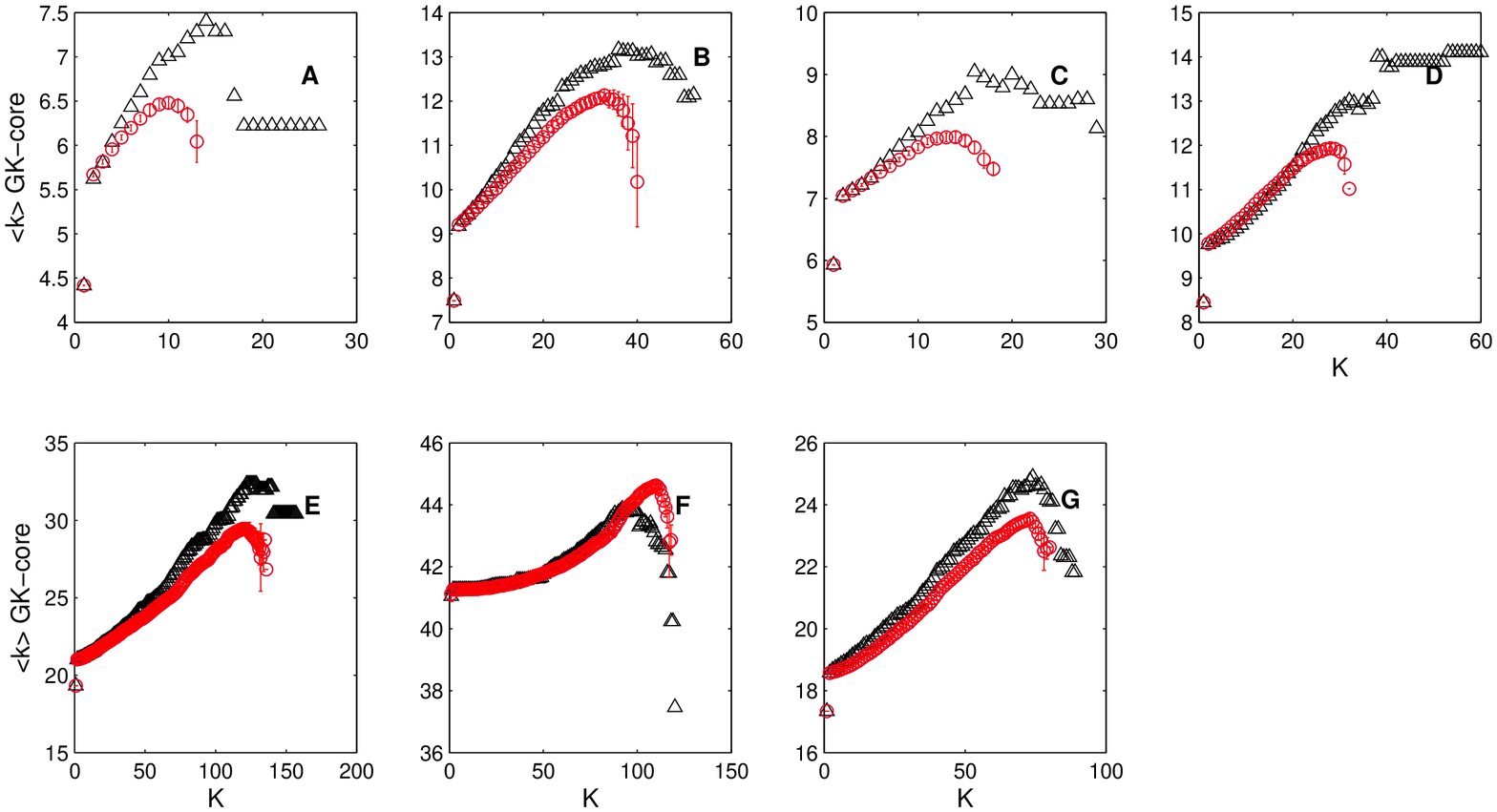}
\caption{Evolution of the average degree of the Giant Connected Component of the $GK$-core for the networks corresponding to the period 796--856 as a function of $K$. a) ${\cal G}_{FTC}$, b) ${\cal G}_{FC}$, c) ${\cal G}_{FT}$, d) ${\cal G}_{TC}$, e) ${\cal G}_C$, f) ${\cal G}_T$, g) ${\cal G}_{F}$. Black triangles depict the behaviour of real networks, red circles and their associated error bars depict the average behaviour of an ensemble of 25 randomised versions of the original networks.}
\end{center}
\end{figure*}

\begin{figure*}
\begin{center}
\includegraphics[width=16.0cm]{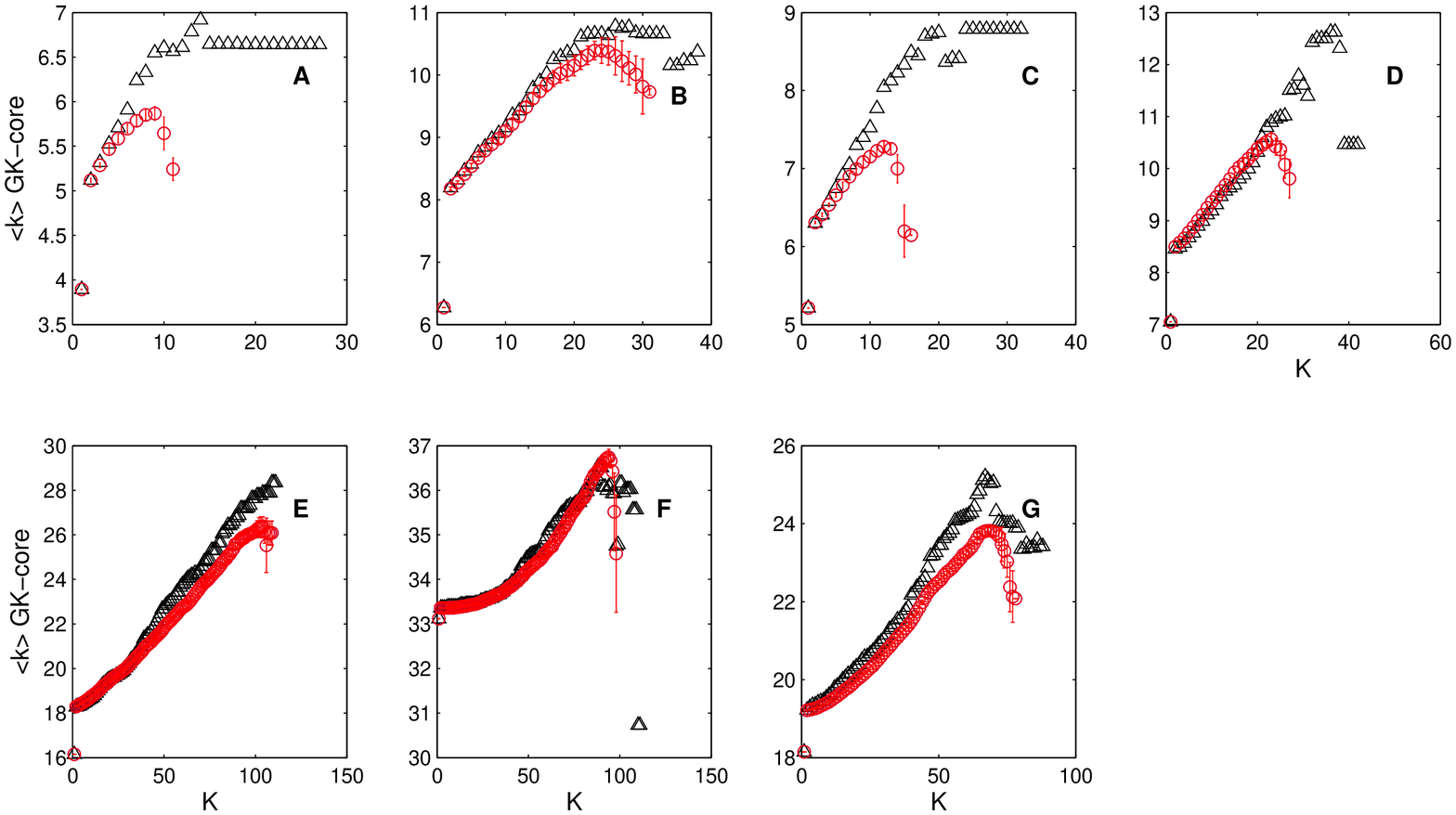}
\caption{Evolution of the average degree of the Giant Connected Component of the $GK$-core for the networks corresponding to the period 1140--1200 as a function of $K$. a) ${\cal G}_{FTC}$, b) ${\cal G}_{FC}$, c) ${\cal G}_{FT}$, d) ${\cal G}_{TC}$, e) ${\cal G}_C$, f) ${\cal G}_T$, g) ${\cal G}_{F}$. Black triangles depict the behaviour of real networks, red circles and their associated error bars depict the average behaviour of an ensemble of 25 randomised versions of the original networks.}
\end{center}
\end{figure*}

\begin{figure*}
\begin{center}
\includegraphics[width=16.5cm]{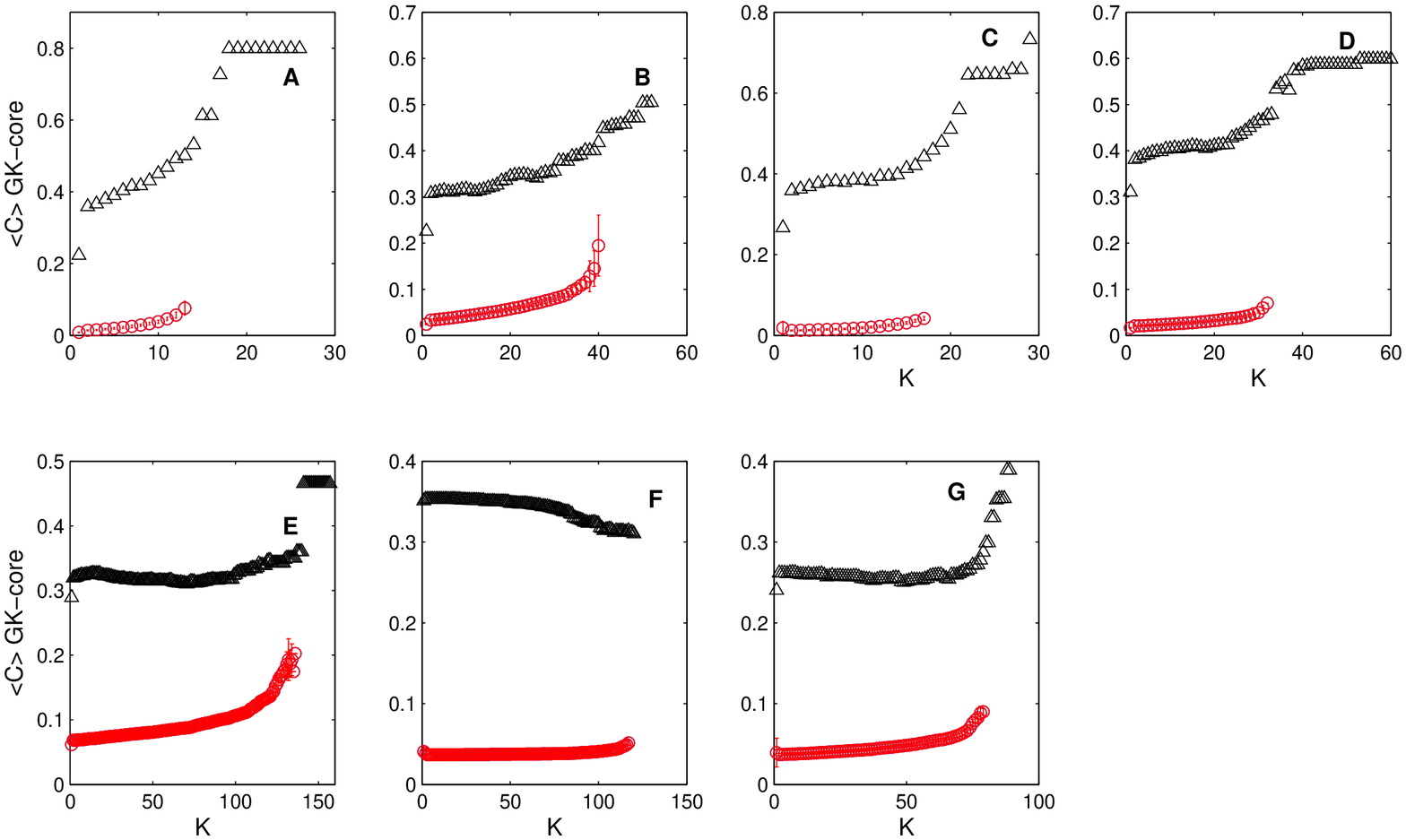}
\caption{Evolution of the average clustering coefficient of the Giant Connected Component of the $GK$-core for the networks corresponding to the period 796--856 as a function of $K$. a) ${\cal G}_{FTC}$, b) ${\cal G}_{FC}$, c) ${\cal G}_{FT}$, d) ${\cal G}_{TC}$, e) ${\cal G}_C$, f) ${\cal G}_T$, g) ${\cal G}_{F}$. Black triangles depict the behaviour of real networks, red circles and their associated error bars depict the average behaviour of an ensemble of 25 randomised versions of the original networks.}
\end{center}
\end{figure*}

\begin{figure*}
\begin{center}
\includegraphics[width=16.5cm]{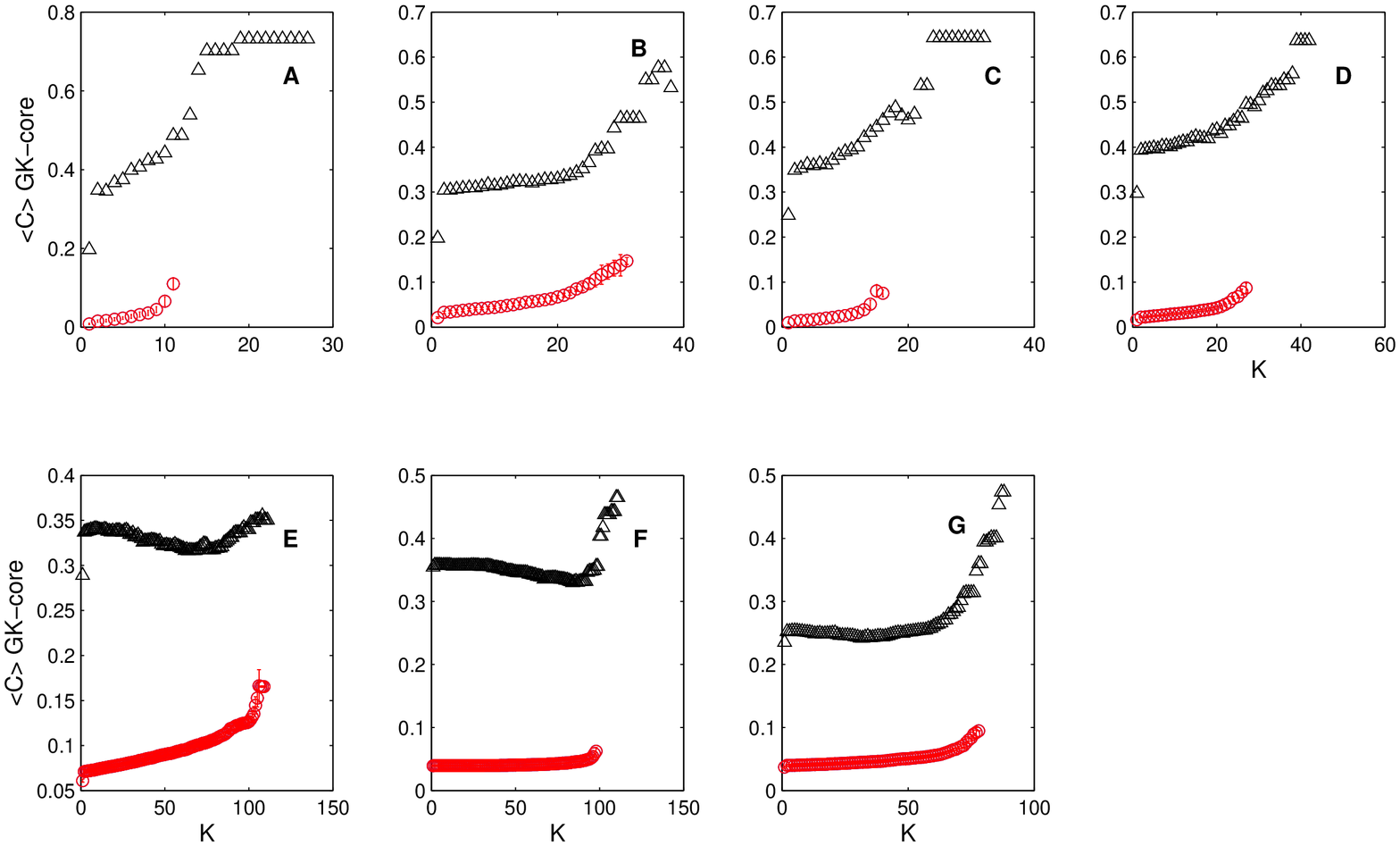}
\caption{Evolution of the average degree of the Giant Connected Component of the $GK$-core for the networks corresponding to the period 1140--1200 as a function of $K$. a) ${\cal G}_{FTC}$, b) ${\cal G}_{FC}$, c) ${\cal G}_{FT}$, d) ${\cal G}_{TC}$, e) ${\cal G}_C$, f) ${\cal G}_T$, g) ${\cal G}_{F}$. Black triangles depict the behaviour of real networks, red circles and their associated error bars depict the average behaviour of an ensemble of 25 randomised versions of the original networks.}
\label{}
\end{center}
\end{figure*}

%\newpage

\begin{table*}[!ht]
\caption{Table with the social indicators. Period 796-856}
ÊÊÊÊ\begin{tabular}{c|c|c|c|c|c|c|c|c}
%%%%%%%
ÊÊ& $\langle$Experience$\rangle$ & $\langle$Activity$\rangle$ & $\langle$Age$\rangle$ & $\langle$Wealth$\rangle$ &  gendComp& FracLead &$\langle$GlobalLead$\rangle$ & NInd  \\ 
\hline
\hline
${\cal G}_{FCT}$ &&&&&&&&\\
\hline 
\hline
Characteristic $GK$ & $4.9\times 10^5$ & $3.6\times 10^6$ & $684$ & $5.37\times 10^7$ & $0.809$ & $0.127$ & $8.53$ & $110$ \\
Hubs & $6.03\times 10^5$ & $4.57\times 10^6$ & $735$ & $5.89\times 10^7$ & $0.947$ & $0.526$ & $9.32$ & $19 $\\
Deepest $GK$ & $6.24\times 10^5$ & $4.45\times 10^6$ & $777$ & $7.89\times 10^7$ & $0.88$ & $0.08$ & $10$ & $25$ \\
Hubs & $1.56\times 10^6$ & $6.14\times 10^6$ & $855$ & $8.03\times 10^7$ & $1$ & $1$ & $13$ & $2 $\\
Deepest $K$-Core 1 & $4.34\times 10^5$ & $3.7\times 10^6$ & $675$ & $4.88\times 10^7$ & $0.7$ & $0.3$ & $9.1$ & $10$ \\
All Net & $3.59\times 10^5$ & $2.81\times 10^6$ & $615$ & $3.19\times 10^7$ & $0.871$ & $0.165$ & $6.84$ & $1564$ \\
\hline
\hline
${\cal G}_{FC}$ &&&&&&&&\\
\hline 
\hline
Critical $GK$ & $3.77\times 10^5$ & $2.97\times 10^6$ & $639$ & $3.45\times 10^7$ & $0.875$ & $0.14$ & $6.96$ & $784 $\\
Hubs & $5.51\times 10^5$ & $3.8\times 10^6$ & $690$ & $4\times 10^7$ & $0.885$ & $0.33$ & $8.24$ & $288$ \\
Deepest $GK$ & $5.71\times 10^5$ & $3.71\times 10^6$ & $691$ & $4.64\times 10^7$ & $0.907$ & $0.215$ & $8.78$ & $107$ \\
Hubs& $8.68\times 10^5$ & $4.97\times 10^6$ & $718$ & $4.26\times 10^7$ & $1$ & $0.8$ & $9.4$ & $5$ \\
Deepest $K$-Core 2 & $7.11\times 10^5$ & $4.36\times 10^6$ & $724$ & $5.16\times 10^7$ & $0.841$ & $0.305$ & $9.18$ & $82$ \\
All Net & $3.43\times 10^5$ & $2.7\times 10^6$ & $610$ & $2.91\times 10^7$ & $0.868$ & $0.145$ & $6.63$ & $1915$ \\
\hline
\hline
%%%%%%%%%%%%%%%%%
${\cal G}_{FT}$&&&&&&&&\\
\hline 
\hline
Characteristic $GK$ & $4.6\times 10^5$ & $3.39\times 10^6$ & $688$ & $4.62\times 10^7$ & $0.797$ & $0.138$ & $8.08$ & $123$ \\
Hubs& $6.54\times 10^5$ & $4.8\times 10^6$ & $792$ & $6.02\times 10^7$ & $0.929$ & $0.429$ & $8.86$ & $14 $\\
Deepest $GK$ & $6.14\times 10^5$ & $4.41\times 10^6$ & $786$ & $7.42\times 10^7$ & $0.857$ & $0.107$ & $10.2$ & $28$ \\
Hubs & $1.56\times 10^6$ & $6.14\times 10^6$ & $855$ & $8.03\times 10^7$ & $1$ & $1$ & $13$ & $2$ \\
Deepest $K$-Core 3 & $4.23\times 10^5$ & $3.71\times 10^6$ & $687$ & $5.53\times 10^7$ & $0.727$ & $0.273$ & $8.91$ & $11$ \\
All Net  & $3.37\times 10^5$ & $2.68\times 10^6$ & $618$ & $2.89\times 10^7$ & $0.871$ & $0.137$ & $6.64$ & $2012$ \\
\hline
\hline
${\cal G}_{CT}$&&&&&&&&\\
\hline 
\hline
Characteristic $GK$ & $3.63\times 10^5$ & $2.8\times 10^6$ & $615$ & $3.02\times 10^7$ & $0.862$ & $0.116$ & $7.06$ & $950$ \\
Hubs & $4.89\times 10^5$ & $3.61\times 10^6$ & $647$ & $5.01\times 10^7$ & $0.905$ & $0.293$ & $8.88$ & $222$ \\
Deepest $GK$ & $6.04\times 10^5$ & $4.04\times 10^6$ & $745$ & $5.93\times 10^7$ & $0.947$ & $0.158$ & $10.1$ & $76$ \\
Hubs & $1.09\times 10^6$ & $5.36\times 10^6$ & $835$ & $9.25\times 10^7$ & $1$ & $0.333$ & $11.3$ & $3$ \\
Deepest $K$-Core 4 & $6.3\times 10^5$ & $4.2\times 10^6$ & $741$ & $7.85\times 10^7$ & $0.971$ & $0.176$ & $10$ & $34 $\\
All Net & $3.18\times 10^5$ & $2.57\times 10^6$ & $606$ & $2.72\times 10^7$ & $0.872$ & $0.128$ & $6.55$ & $2196$ \\
\hline
\hline
${\cal G}_C$&&&&&&&&\\
\hline 
\hline
Deepest $GK$ & $5.37\times 10^5$ & $3.68\times 10^6$ & $693$ & $4.22\times 10^7$ & $0.915$ & $0.23$ & $8.15$ & $248$ \\
Hubs & $9.71\times 10^5$ & $5.14\times 10^6$ & $709$ & $5.18\times 10^7$ & $1$ & $0.5$ & $8.5$ & $4$ \\
Deepest $K$-Core  & $6.67\times 10^5$ & $4.24\times 10^6$ & $716$ & $4.63\times 10^7$ & $0.93$ & $0.279$ & $8.76$ & $129$ \\
\hline
\hline
${\cal G}_T$&&&&&&&&\\
\hline 
\hline
Deepest $GK$ & $4.27\times 10^5$ & $3.11\times 10^6$ & $657$ & $3.45\times 10^7$ & $0.884$ & $0.149$ & $7$ & $1019$ \\
Hubs & $5.17\times 10^5$ & $4.15\times 10^6$ & $766$ & $6.69\times 10^7$ & $0.733$ & $0.167$ & $9.1$ & $30$ \\
Deepest $K$-Core & $3.82\times 10^5$ & $2.88\times 10^6$ & $600$ & $3.34\times 10^7$ & $0.896$ & $0.127$ & $8.49$ & $347$ \\
\hline
\hline
${\cal G}_F$&&&&&&&&\\
\hline 
\hline
Characteristic $GK$ & $1.72\times 10^5$ & $1.66\times 10^6$ & $512$ & $1.52\times 10^7$ & $0.874$ & $0.0402$ & $5.69$ & $697$ \\
Hubs & $4.07\times 10^5$ & $3.13\times 10^6$ & $687$ & $3.35\times 10^7$ & $0.87$ & $0.171$ & $6.96$ & $1487 $\\
Deepest $GK$  & $5.34\times 10^5$ & $3.6\times 10^6$ & $712$ & $4.3\times 10^7$ & $0.868$ & $0.196$ & $7.31$ & $372$ \\
Hubs & $9.32\times 10^5$ & $4.96\times 10^6$ & $807$ & $6.18\times 10^7$ & $0.818$ & $0.455$ & $10.1$ & $11$ \\
Deepest $K$-Core  & $6.81\times 10^5$ & $4.33\times 10^6$ & $775$ & $4.85\times 10^7$ & $0.881$ & $0.252$ & $8.1$ & $218$ \\
%%%%%%%%%%%%%%%%%
\hline
\hline
All Players & $3.07\times 10^5$ & $2.5\times 10^6$ & $606$ & $2.59\times 10^7$ & $0.87$ & $0.119$ & $6.39$ & $2422 $\\
%\hline
ÊÊÊ\end{tabular}
\end{table*}
\begin{table*}[!ht]
\caption{Table with the social indicators. Period 1140-1200}
ÊÊÊÊ\begin{tabular}{c|c|c|c|c|c|c|c|c}
%%%%%%%
ÊÊ& $\langle$Experience$\rangle$ & $\langle$Activity$\rangle$ & $\langle$Age$\rangle$ & $\langle$Wealth$\rangle$ & $ gendComp$ & FracLead &$\langle$GlobalLead$\rangle$ & NInd  \\ 
\hline
\hline
${\cal G}_{FCT}$ &&&&&&&&\\
\hline 
\hline
Characteristic $GK$  & $7.72\times 10^5$ & $5.69\times 10^6$ & $1.02\times 10^3$ & $9.84\times 10^7$ & $0.885$ & $0.195$ & $10.7$ & $87$ \\
Hubs& $1.01\times 10^6$ & $6.86\times 10^6$ & $1.08\times 10^3$ & $1.23\times 10^8$ & $0.933$ & $0.4$ & $11.4$ & $15 $\\
Deepest $GK$ & $9.78\times 10^5$ & $5.96\times 10^6$ & $1.09\times 10^3$ & $1.14\times 10^8$ & $0.962$ & $0.154$ & $11.3$ & $26$ \\
Hubs & $5.69\times 10^5$ & $7.39\times 10^6$ & $1.2\times 10^3$ & $3.03\times 10^8$ & $1$ & $1$ & $12$ & $2$ \\
Deepest $K$-Core  & $7.18\times 10^5$ & $6.23\times 10^6$ & $1.09\times 10^3$ & $1.4\times 10^8$ & $0.889$ & $0.111$ & $11$ & $9$ \\
All Net & $4.86\times 10^5$ & $3.88\times 10^6$ & $857$ & $4.87\times 10^7$ & $0.875$ & $0.165$ & $7.64$ & $1303$ \\
\hline
\hline
${\cal G}_{FC}$ &&&&&&&&\\
\hline 
\hline
Characteristic $GK$ & $8.47\times 10^5$ & $5.72\times 10^6$ & $1.04\times 10^3$ & $7.69\times 10^7$ & $0.884$ & $0.207$ & $9.41$ & $121$ \\
HUBS & $1.32\times 10^6$ & $6.96\times 10^6$ & $1.15\times 10^3$ & $1.24\times 10^8$ & $0.778$ & $0.333$ & $12.6$ & $9$ \\
Deepest $GK$& $8.07\times 10^5$ & $5.59\times 10^6$ & $1.01\times 10^3$ & $6.37\times 10^7$ & $0.882$ & $0.235$ & $8.69$ & $85$ \\
Hubs & $1.53\times 10^6$ & $6.84\times 10^6$ & $1.13\times 10^3$ & $7.26\times 10^7$ & $0.714$ & $0.143$ & $12.7$ & $7$ \\
Deepest $K$-Core & $9.4\times 10^5$ & $6.03\times 10^6$ & $1.01\times 10^3$ & $6.66\times 10^7$ & $0.882$ & $0.329$ & $9.5$ & $76$ \\
All Net & $4.69\times 10^5$ & $3.72\times 10^6$ & $842$ & $4.35\times 10^7$ & $0.871$ & $0.154$ & $7.4$ & $1600 $\\
\hline
\hline
%%%%%%%%%%%%%%%%%
${\cal G}_{FT}$&&&&&&&&\\
\hline 
\hline
Characteristic $GK$ & $8.48\times 10^5$& $5.77\times 10^6$ & $1.05\times 10^3$ & $8.94\times 10^7$ & $0.892$ & $0.169$ & $10.6$ & $83 $\\
Hubs & $1.34\times 10^6$ & $7.37\times 10^6$ & $1.13\times 10^3$ & $1.8\times 10^8$ & $0.889$ & $0.333$ & $12.1$ & $9$ \\
Deepest $GK$  & $9.2\times 10^5$ & $5.87\times 10^6$ & $1.11\times 10^3$ & $1.1\times 10^8$ & $0.935$ & $0.194$ & $11.3$ & $31$ \\
Hubs & $5.69\times 10^5$ & $7.39\times 10^6$ & $1.2\times 10^3$ & $3.03\times 10^8$ & $1$ & $1$ & $12$ & $2$ \\
Deepest $K$-Core & $7.18\times 10^5$ & $6.23\times 10^6$ & $1.09\times 10^3$ & $1.4\times 10^8$ & $0.889$ & $0.111$ & $11$ & $9$ \\
All Net & $4.76\times 10^5$ & $3.77\times 10^6$ & $869$ & $4.46\times 10^7$ & $0.872$ & $0.143$ & $7.56$ & $1660$ \\
\hline
\hline
${\cal G}_{CT}$&&&&&&&&\\
\hline 
\hline
Characteristic $GK$ & $7.38\times 10^5$ & $5.34\times 10^6$ & $989$ & $9.17\times 10^7$ & $0.934$ & $0.231$ & $10.9$ & $91$ \\
Hubs & $4.68\times 10^5$ & $6.88\times 10^6$ & $1.11\times 10^3$ & $1.7\times 10^8$ & $1$ & $0.6$ & $11.8$ & $5$ \\
Deepest $GK$ & $7.06\times 10^5$ & $5.43\times 10^6$ & $1.02\times 10^3$ & $1\times 10^8$ & $0.927$ & $0.341$ & $10.7$ & $41$ \\
Hubs & $2.98\times 10^5$ & $5.87\times 10^6$ & $982$ & $4.03\times 10^7$ & $1$ & $0.5$ & $11$ & $2 $\\
Deepest $K$-Core  & $9.53\times 10^5$ & $5.99\times 10^6$ & $1.03\times 10^3$ & $1.1\times 10^8$ & $0.912$ & $0.206$ & $11.3$ & $34$ \\
All Net & $4.33\times 10^5$ & $3.54\times 10^6$ & $831$ & $4.22\times 10^7$ & $0.871$ & $0.137$ & $7.44$ & $1788 $\\
\hline
\hline
${\cal G}_C$&&&&&&&&\\
\hline 
\hline
Deepest $GK$ & $6.25\times 10^5$ & $4.49\times 10^6$ & $884$ & $5.53\times 10^7$ & $0.888$ & $0.24$ & $8.21$ & $483$ \\
Hubs 5 & $1.21\times 10^6$ & $6.74\times 10^6$ & $1.03\times 10^3$ & $6.25\times 10^7$ & $0.929$ & $0.429$ & $10.1$ & $14$ \\
Deepest $K$-Core  & $8.23\times 10^5$ & $5.57\times 10^6$ & $968$ & $7.53\times 10^7$ & $0.874$ & $0.394$ & $8.77$ & $127$ \\
\hline
\hline
${\cal G}_T$&&&&&&&&\\
\hline 
\hline
Deepest $GK$ & $4.27\times 10^5$ & $3.11\times 10^6$ & $657$ & $3.45\times 10^7$ & $0.884$ & $0.149$ & $7$ & $1019$ \\
Hubs & $5.17\times 10^5$ & $4.15\times 10^6$ & $766$ & $6.69\times 10^7$ & $0.733$ & $0.167$ & $9.1$ & $30 $\\
Deepest $K$-Core  & $3.82\times 10^5$ & $2.88\times 10^6$ & $600$ & $3.34\times 10^7$ & $0.896$ & $0.127$ & $8.49$ & $347 $\\
\hline
\hline
${\cal G}_F$&&&&&&&&\\
\hline 
\hline
Characteristic $GK$ & $1.9\times 10^5$ & $1.88\times 10^6$ & $608$ & $1.86\times 10^7$ & $0.86$ & $0.0457$ & $6.08$ & $328$ \\
Hubs & $5.05\times 10^5$ & $4\times 10^6$ & $925$ & $4.65\times 10^7$ & $0.87$ & $0.155$ & $7.61$ & $1585$ \\
Deepest $GK$ & $7.57\times 10^5$ & $5.34\times 10^6$ & $1.05\times 10^3$ & $5.96\times 10^7$ & $0.877$ & $0.175$ & $7.33$ & $171$ \\
Hubs & $1.39\times 10^6$ & $6.68\times 10^6$ & $1.15\times 10^3$ & $7.8\times 10^7$ & $0.6$ & $0$ & $12.8$ & $5 $\\
Deepest $K$-Core  & $1\times 10^6$ & $6.12\times 10^6$ & $1.08\times 10^3$ & $6.83\times 10^7$ & $0.88$ & $0.253$ & $9.11$ & $83$ \\
%All Net KCore 7 & $4.29\times 10^5$ & $3.47\times 10^6$ & $841$ & $3.95\times 10^7$ & $0.869$ & $0.127$ & $7.23$ & $2058$ \\
%%%%%%%%%%%%%%%%%
\hline
\hline
${\rm All}\;{\rm players}$ 	& $4.3\times 10^5$ 	& $3.5 \times 10^6$ 	& $841$	& $3.96\times 10^7$  &$0.87$    &  $0.12$    & $7.51$   & $2059$
%\hline
ÊÊÊ\end{tabular}
\end{table*}

\end{document}